# Kinematics of Abdominal Aortic Aneurysms

Mostafa Jamshidian*, Adam Wittek, Saeideh Sekhavat, Karol Miller

Intelligent Systems for Medicine Laboratory, The University of Western Australia, Perth, Western Australia, Australia

A search in Scopus within "Article title, Abstract, Keywords" unveils 2,444 documents focused on the biomechanics of Abdominal Aortic Aneurysm (AAA), mostly on AAA wall stress. Only 24 documents investigated AAA kinematics, an important topic that could potentially offer significant insights into the biomechanics of AAA. In this paper, we present an image-based approach for patient-specific, in vivo, and non-invasive AAA kinematic analysis using patient's time-resolved 3D computed tomography angiography (4D-CTA) images, with an objective to measure wall displacement and strain during the cardiac cycle. Our approach relies on regularized deformable image registration for estimating wall displacement, estimation of the local wall strain as the ratio of its normal displacement to its local radius of curvature, and local surface fitting with non-deterministic outlier detection for estimating the wall radius of curvature. We verified our approach against synthetic ground truth image data created by warping a 3D-CTA image of AAA using a realistic displacement field obtained from a finite element biomechanical model. We applied our approach to assess AAA wall displacements and strains in ten patients. Our kinematic analysis results indicated that the 99th percentile of circumferential wall strain, among all patients, ranged from 2.62% to 5.54%, with an average of 4.45% and a standard deviation of 0.87%. We also observed that AAA wall strains are significantly lower than those of a healthy aorta. Our work demonstrates that the registration-based measurement of AAA wall displacements in the direction normal to the wall is sufficiently accurate to reliably estimate strain from these displacements.

---

* Corresponding author.
E-mail address: mostafa.jamshidian@uwa.edu.au (M. Jamshidian).

## 1. Introduction

Abdominal aortic aneurysm (AAA) is a permanent and irreversible dilation of the lower aorta that is often asymptomatic, and diagnosis frequently occurs via incidental findings. While asymptomatic, AAA can expand to the point of rupture if left untreated, leading to death in most cases (NICE, 2020; Wanhainen et al., 2019).

The current clinical practice for AAA management, based on the maximum diameter and its growth rate is a one-size-fits-all approach that recommends clinical intervention when the aneurysm diameter exceeds 5.5 cm in men and 5 cm in women, or when growth rates exceed 1 cm per year (Wanhainen et al., 2019). The maximum diameter criteria may underestimate or overestimate the rupture risk in individual AAA patients, as evidenced by ruptured AAAs with diameters smaller than the critical diameter (Vorp, 2007) and unruptured stable AAAs with diameters larger than the critical diameter (Darling et al., 1977; NICE, 2020; Wanhainen et al., 2019). Autopsy findings indicate that around 13% of AAAs with a maximum diameter of 5 cm or less experienced rupture, whereas 60% of AAAs larger than 5 cm in diameter remained unruptured (Kontopodis et al., 2016).

AAA biomechanics, particularly wall stress, has been extensively studied to customize disease management to individual patients (Chung et al., 2022; Farotto et al., 2018; Fillinger et al., 2003; Fillinger et al., 2002; Gasser et al., 2010; Gasser et al., 2022; Indrakusuma et al., 2016; Joldes et al., 2016; Joldes et al., 2017; Li et al., 2008; Liddelow et al., 2023; Miller et al., 2020; Polzer et al., 2020; Singh et al., 2023; Speelman et al., 2007; Vande Geest et al., 2006; Wang et al., 2023). Despite advancements in biomechanical models that compute patient-specific AAA wall stress without requiring the AAA wall mechanical properties and thickness distribution (Joldes et al., 2016), stress-based rupture risk indicators still require information about wall strength (Singh et al., 2023).

Direct non-invasive measurement of patient-specific in-vivo wall strength (or degradation) is currently impossible, but it may be indirectly estimated by combining the wall stress and strain maps (Nagy et al., 2015). Only a few research groups explored non-invasive, in-vivo measurement of AAA wall displacement and strain using sequential images at different phases in the cardiac cycle, often referred to as 4D images (Cebull et al., 2019; Derwich et al., 2023;

Maas et al., 2024; Nagy et al., 2015; Raut et al., 2014; Satriano et al., 2015; Wang et al., 2018; Wittek et al., 2017; Wittek et al., 2018). Raut et al. (2014) extracted the systolic and diastolic AAA geometries from MRI images using an in-house framework for image-based reconstruction of tessellated surfaces to evaluate wall displacement and strain. Wittek et al. (2017) and Maas et al. (2024) combined speckle tracking algorithms with 4D ultrasound for measuring displacement and strain in AAA. Nagy et al. (2015) estimated AAA kinematic fields with an isogeometric shell analysis framework with smoothing to mitigate measurement inaccuracy. Based on fluid-structure interaction simulation, they assumed that the incremental displacement vector is normal to the AAA surface. Satriano et al. (2015) used optical flow of time-wise consecutive images to first compute deformation and then aortic wall distensibility or deformability.

In this paper, we developed, verified, and applied to ten real AAAs, a patient-specific approach to AAA kinematic analysis using 4D-CTA image data. We used regularized deformable image registration to estimate wall displacement, calculated local wall strain as the ratio of normal displacement to the local radius of curvature, and employed local surface fitting with non-deterministic outlier detection to determine the wall radius of curvature.

The remainder of the paper is organized as follows: In Section 2, we presented the image data of AAA patients and the methods for estimating wall displacement and strain from registration, as well as the preparation of synthetic ground truth image data. In Section 3, We verified our approach against synthetic ground truth. In Section 4, we presented AAA wall displacements and strains in ten patients, followed by our conclusions and discussions in Section 5.

## 2. Materials and methods

### 2.1. Image data

We used anonymized contrast-enhanced 4D-CTA image datasets, each consisting of ten 3D volume frames per cardiac cycle, from ten patients diagnosed with AAA (See Supplementary Material, Section S1, for details). As an example, Figure 1a shows the cropped 3D-CTA image of Patient 1's AAA in the diastolic phase. This image has dimensions of $(112, 109, 74)$ voxels and voxel spacing of $(0.63, 0.63, 1.00)$ mm along $(\mathrm{R}, \mathrm{A}, \mathrm{S})$ axes, as illustrated in Figure 1a. In the patient coordinate system, the basis is aligned with the anatomical axes of anterior-posterior, inferior-superior, and left-right. Specifically, the axes R, A, and S correspond, respectively, to left-right, posterior-anterior, and inferior-superior directions.

### 2.2. Image registration, displacement, and strain

In this study, we used deformable image registration to align the systolic and diastolic 3D frames of 4D-CTA and estimate the displacement field that maps systolic to diastolic AAA geometry. We selected these frames because the wall displacements between them are the largest, resulting in the greatest strains, which can be extracted most reliably. We then extracted wall displacements from the registration displacement map and subsequently computed strain.

*Regularized image registration theory:* We used MATLAB implementation of a deformable image registration with isotropic total variation regularization of displacement (Vishnevskiy et al., 2017). The regularized image registration, within a discretized image domain $\Omega$, estimates the three-dimensional displacement field $\boldsymbol{d} : \Omega \to \mathbb{R}$ that maps a moving image $\boldsymbol{I}_M$ onto a fixed image $\boldsymbol{I}_F$ via the following optimization problem:

$$\boldsymbol{d}^{\star} = \operatorname*{argmin}_{\boldsymbol{d}} \boldsymbol{\mathcal{F}}(\boldsymbol{d}) = \operatorname*{argmin}_{\boldsymbol{d}} E_D(\boldsymbol{d}; \boldsymbol{I}_F, \boldsymbol{I}_M) + \lambda E_R(\boldsymbol{d}), \quad (1)$$

where $E_D$ is an image dissimilarity metric and $E_R$ is the displacement regularization term with coefficient $\lambda$ controlling the amount of regularization. The fixed and moving images are composed of $L = |\Omega|$ voxels. Each voxel has the physical dimensions $(\delta_1, \delta_2, \delta_3)$ along directions $(1, 2, 3)$ and the volume $v = \delta_1 \delta_2 \delta_3$.

For subsequent analysis, we treated the images and the displacement components fields as vectors i.e., $\boldsymbol{I} \in \mathbb{R}^{\mathrm{L}}$ and $\boldsymbol{d}_i \in \mathbb{R}^{\mathrm{L}}$ for $i = 1, 2, 3$ with $\boldsymbol{d}_i$ being the $i$th component of the displacement field $\boldsymbol{d} = (\boldsymbol{d}_1^{\top}, \boldsymbol{d}_2^{\top}, \boldsymbol{d}_3^{\top})^{\top} \in \mathbb{R}^{L\times 3}$.

To reduce the dimensionality of search space in the optimization problem (Eq. 1), we parametrized the displacement field $\boldsymbol{d}$ by interpolating the displacement field $\boldsymbol{k}$ on a structured and evenly spaced grid of $M$ control points i.e., $\boldsymbol{d} = \boldsymbol{d}(\boldsymbol{k})$. The control point grid spacing is $(K_1, K_2, K_3)$ voxels or in physical dimensions $(\delta_1 K_1, \delta_2 K_2, \delta_3 K_3)$, along directions $(1, 2, 3)$. Then, the cell volume of the control point grid is $\eta = vK_1K_2K_3 = \delta_1\delta_2\delta_3K_1K_2K_3$.

For subsequent analysis, we treated the displacement components of control points as vectors i.e., $\boldsymbol{k}_i \in \mathbb{R}^M$ for $i = 1, 2, 3$ with $\boldsymbol{k}_i$ being the $i$th component of the control points displacement field $\boldsymbol{k} = (\boldsymbol{k}_1^\top, \boldsymbol{k}_2^\top, \boldsymbol{k}_3^\top)^\top \in \mathbb{R}^{M \times 3}$.

While cubic B-splines are more common in deformable image registration, in this study, we used first-order B-splines with smaller spatial support to ensure that the interpolated displacements are bounded by the control point displacements and to eliminate overshooting effects when approximating sharp image gradients, such as those in the AAA wall region.

As a smooth image dissimilarity metric, we employed the local correlation coefficient (LCC) between the two images $\boldsymbol{I}_F$ and $\boldsymbol{I}_M$ defined as (Cachier and Pennec, 2000):

$$\mathrm{LCC}(\boldsymbol{I}_F, \boldsymbol{I}_M) = -\sum_{\boldsymbol{x} \in \Omega} \frac{\overline{\boldsymbol{I}_F \circ \boldsymbol{I}_M} - \overline{\boldsymbol{I}_F} \circ \overline{\boldsymbol{I}_M}}{\sqrt{\overline{\boldsymbol{I}_F^2} - \overline{\boldsymbol{I}_F}^2} \circ \sqrt{\overline{\boldsymbol{I}_M^2} - \overline{\boldsymbol{I}_M}^2}} [\boldsymbol{x}] v, \tag{2}$$

with $\circ$ denoting the elementwise multiplication operator. The image average $\overline{\boldsymbol{I}} = H_w \otimes \boldsymbol{I}$ is calculated by convolutions using a Gaussian weighting kernel $H_w$ with standard bandwidth $w$, with $\otimes$ denoting the convolution operator. Also, the operator $\boldsymbol{I}[\boldsymbol{x}]$ interpolates the image $\boldsymbol{I}$ at position $\boldsymbol{x}$.

We imposed total variation regularization on the control point displacement field $\boldsymbol{k}$ rather than the dense voxel displacement field $\boldsymbol{d}$, as follows:

$$E_R^{TV}(\boldsymbol{k}) = \eta \sum_{m \leq M} \sqrt{\sum_{i,j \leq 3} \left(\nabla_i k_j[m]\right)^2}, \tag{3}$$

where $\nabla_i$ denotes the derivative operator along the $i$th direction, and $k_j[m]$ for $j = 1, 2, 3$ denotes the $j$th component of the control point displacement field at the control point index $m$. Hence, $\boldsymbol{k}[m] = (k_1[m], k_2[m], k_3[m])^\top \in \mathbb{R}^3$ represents the displacement vector of the $m$th control point. Total variation regularization is more effective in estimating displacements that do not align with the Cartesian coordinate axes (Vishnevskiy et al., 2017).

Finally, we reformulated the optimization problem for the total variation-regularized image registration in terms of the control point displacement field, as follows:

$$\boldsymbol{k}^\star = \underset{\boldsymbol{k}}{\operatorname{argmin}}\, \boldsymbol{\mathcal{F}}\big(\boldsymbol{d}(\boldsymbol{k})\big) = \underset{\boldsymbol{d}}{\operatorname{argmin}}\, E_D(\boldsymbol{d}(\boldsymbol{k}); \boldsymbol{I}_F, \boldsymbol{I}_M) + \lambda E_R^{TV}(\boldsymbol{k}). \tag{4}$$

For more details on the image registration algorithm, see Supplementary Material, Section S2.

*Wall displacement:* Registering the undeformed image onto the deformed image yielded the control point displacement field $\boldsymbol{k}$ that was then interpolated at the AAA wall location to obtain the wall displacement from registration in the Cartesian patient coordinate system $(\mathrm{R}, \mathrm{A}, \mathrm{S})$.

Registration is known to produce more accurate displacements in the direction of image gradients (Lehoucq et al., 2021). For AAA 3D-CTA images, the image gradient in the wall region mostly aligns with the wall normal. Therefore, for subsequent analysis, we established a local biological coordinate system comprising the local normal and two perpendicular local tangents to the wall surface.

To extract a smooth field of unit normal vectors on the AAA surface, we employed the planar least squares regression method in which the normal at each point in the point cloud is determined by fitting a local plane using neighbouring points (See Supplementary Material, Section S3, for details). For AAA, the point cloud consists of the STL vertices on the external surface of the AAA wall model. Figure 2 shows the unit normal vectors on AAA surface, obtained via planar regression, which is a smooth vector field suitable for defining biological coordinates.

*Wall strain:* As will be discussed later in Section 3, the component-wise analysis of the registration displacement field revealed that the tangential displacement from registration is an unreliable measurement. Furthermore, although the normal displacement from registration was satisfactorily estimated, it was noisy, and its spatial derivatives for strain calculation would intensify this noise. Therefore, any differential kinematic description of wall deformation must be based solely on the normal registration displacement.

Wittek et al. (2018) employed finite element method to compute strain field using wall motion data from 4D ultrasound and demonstrated the predominance of circumferential strain as a physiologically meaningful measurement of AAA wall strain. Karatolios et al. (2013) demonstrated that the mean of this circumferential strain is in good agreement with the ratio of diameter change to diameter, which represents the circumferential strain of a uniformly inflated membrane with circular cross-section (Coelho et al., 2014; Fichter, 1997; Hencky, 1915). Similarly, Morrison et al. (2009) calculated the Green strain component in the circumferential direction as $0.5(\lambda^2 - 1)$, where $\lambda$ represented the systolic to diastolic circumference ratio, equivalent to the ratio of systolic to diastolic mean diameters. Guided by (Karatolios et al., 2013; Morrison et al., 2009), we adopted the following localized measure of circumferential wall strain:

$$\epsilon = \frac{\Delta r}{r}, \tag{5}$$

where $r$ is the local radius of wall curvature and $\Delta r$ is the local normal displacement of the wall obtained via registration. Under the assumption that the aneurysm wall locally deforms as a uniformly expanding (with no axial stretching) homogenous cylinder made of incompressible material, our simplified measure of strain (Eq. 6) corresponds to the following Green strain tensor in the local cylindrical coordinate system composed of radial, circumferential, and axial directions, respectively:

$$\boldsymbol{E} = \begin{bmatrix} \frac{1}{2(\epsilon+1)^2} - \frac{1}{2} & 0 & 0 \\ 0 & \frac{(\epsilon+1)^2}{2} - \frac{1}{2} & 0 \\ 0 & 0 & 0 \end{bmatrix}. \tag{6}$$

To estimate the local radius of curvature for the wall point cloud, we utilized a variant of Random Sample Consensus (RANSAC) algorithm known as the M-estimator SAmple Consensus (MSAC) algorithm (Torr and Zisserman, 2000), available in MATLAB. We modified the native MATLAB algorithm to fit a cylinder to a local neighbourhood of points around a typical query point in the wall point cloud. We adopted the radius of the fitted cylinder as the local radius of curvature of the wall at that query point. The local neighbourhood of points consists of the K-nearest neighbours of the query point, determined by the k-d tree based search algorithm (Muja and Lowe, 2009), available in MATLAB. To achieve a reliable curvature measurement, we applied an orientation constraint to align the fitted cylinder axis with a reference orientation vector chosen as the inferior-superior or S axis for AAA (see Figure 1 for S axis definition).

Figure 3 shows the contour plots for radius of curvature for the external surface of AAA wall. The vector plots in Figure 3 illustrate that the modified MSAC algorithm effectively constrains the orientation of the fitted cylinder to align with the S axis.

### 2.3. Synthetic ground truth

For verification of our method, we created synthetic ground truth image data by warping Patient 1's 3D-CTA image using a realistic wall displacement obtained from a patient-specific finite element biomechanical model. In accordance with FDA (Food and Drug Administration, 2023) and ASME (The American Society of Mechanical Engineers, 2018) guidelines, the plausibility of the synthetic ground truth, rather than exact biomechanical fidelity, is sufficient for this process.

*Biomechanical model:* We used PRAEVAorta by NUREA (https://www.nurea-soft.com/) for AI-based automatic segmentation of Patient 1's AAA image (Figure 1a), followed by automated post-processing with in-house MATLAB code (Alkhatib et al., 2024), to automatically extract the voxelated geometries of AAA wall and blood. We used these geometries and the assumed wall thickness of 1.5 mm (Raut et al., 2013) as input for the BioPARR (Biomechanics-based Prediction of Aneurysm Rupture Risk) software package (Joldes, 2024; Joldes et al., 2017) to automatically create AAA model, as shown in Figure 1b. We used hexahedral elements to mesh the model in HyperMesh (Alkhatib et al., 2023a; Altair, 2024). We imported the mesh into the Abaqus finite element software (Simulia, 2024). The mesh, shown in Figure 1c, consists of 49530 nodes and 8892 20-node quadratic hexahedral elements with hybrid formulation and constant pressure (element type C3D20H in Abaqus) (Alkhatib et al., 2023b). We fixed the top and bottom ends of AAA and applied a uniform pressure of 13 kPa on the inner surface of the wall. We used a hyperelastic material model for the wall tissue material, with the polynomial strain energy potential given by (Raghavan and Vorp, 2000):

$$W = \alpha(I_B - 3) - \beta(I_B - 3)^2, \tag{7}$$

where $I_B$ is the first invariant of the left Cauchy–Green or Finger deformation tensor $\boldsymbol{B}$. The material parameters $\alpha = 0.174 \text{ MPa}$ and $\beta = 1.881 \text{ MPa}$ are from the literature (Fillinger et al., 2002; Raghavan and Vorp, 2000; Raghavan et al., 2000). Finally, we solved Patient 1's AAA finite element model using Abaqus/Standard solver (Simulia, 2024).

Figure 4 displays the synthetic ground truth AAA wall deformation calculated using the AAA biomechanical model shown in Figure 1c. This figure highlights a non-uniform distribution of wall displacement, with a maximum of 1.37 mm.

*Verification approach and synthetic systolic image:* We warped Patient 1's AAA image in the diastolic phase, shown in Figure 1a, using 3D Slicer Transforms module to create the synthetic systolic image (Fedorov et al., 2012; Slicer, 2024). We used the Scattered Transform Extension (Joldes, 2017) in 3D Slicer to create a B-Spline warping transform based on the biomechanics-calculated AAA wall displacement as ground truth. The Scattered Transform was originally developed to warp preoperative brain images based on a deformation field obtained from a biomechanical model (Joldes et al., 2012).

## 3. Results

### 3.1. Method verification

*Image registration:* To estimate wall displacement from registration, we employed Patient 1's diastolic (undeformed) and synthetic systolic (deformed) images as the moving and fixed images, respectively, in deformable image registration. We set the control points spacing $(K_1, K_2, K_3) = (6, 6, 6)$ voxels, and the regularization coefficient $\lambda = 0.05$.

Figure 5 compares the moving versus fixed and the registered versus fixed images by their Canny edge overlays (Canny, 1986). While the Canny edge overlays demonstrate a good match between the fixed and registered images, we carefully examined the registration displacement field, as follows, before strain calculation.

*Wall displacement:* With the local biological coordinates established, as described in Section 2.2, we decomposed the local wall displacement vector into its normal and tangential components. Figure 6 shows the contour plots of displacement magnitude, normal displacement, and tangential displacement for both the registration and ground truth displacement fields, along with the contour plots of the differences between them. Figure 7 provides a closer examination of the displacements through scatter plots depicting the registration versus ground truth displacements for displacement magnitude, normal displacement, and tangential displacement. Each plot in Figure 7 displays the R-squared value and Normalized Root Mean Square Error (NRMSE) for the identity line fit (y = x).

Figure 7 shows that the normal displacement component exhibits a relatively tight scatter around the identity line, with R-squared of 0.981 and NRMSE of 0.028, indicating a better correspondence between registration and ground truth, compared to the tangential displacement component. Noteworthy is that both the registration and ground truth normal displacement vectors align with the surface normal and differ only in magnitude. However, the registration and ground truth tangential displacement vectors differ in both magnitude and direction.

Figure 7 also demonstrates that the tangential displacement component displays a wide scatter around the identity line, with R-squared of 0.841 and NRMSE of 0.072, indicating significant discrepancies between the registration and ground truth tangential displacement magnitude. Furthermore, the colour code in the tangential displacement scatter plot in Figure 7 illustrates large differences in tangential displacement direction between registration and ground truth, particularly pronounced for smaller values of tangential displacement.

Therefore, we discarded the unreliable tangential displacement and used only the normal displacement in strain calculation. The prominence of the normal wall displacement has previously been hypothesized based on observations from fluid-structure interaction simulations (Nagy et al., 2015).

*Wall strain:* Figure 8 compares the wall strain obtained from registration, defined as the ratio of normal registration displacement to the local wall radius of curvature, against the ground truth wall strain. Figures 8a and 8b display the contour plots of registration and ground truth wall strain, respectively, and Figure 8c shows the difference between them. Figure 8d offers a detailed analysis of strain verification through a scatter plot comparing the registration and ground truth strain. The relatively tight scatter in Figure 8d, with R-squared of 0.941 and NRMSE of 0.028 for the identity line fit (y = x), verifies the methods we developed for AAA kinematic analysis.

Due to the features of the methods that we used for estimating wall displacement and curvature, the wall strain may include outliers arising from large displacement outliers and small radius of curvature outliers. Therefore, following the practice used in AAA stress analysis (Alkhatib et al., 2024; Speelman et al., 2008; Wittek et al., 2022), we reported the 99th percentile of strain, rather than the peak strain. The 99th percentile of wall strain from the registration and ground truth were 5.49% and 5.36%, respectively, implying a relative difference of 2.4%. This further verifies our AAA kinematic analysis method (See Supplementary Material, Section S4, for normality check of the data).

### 3.2. AAA Kinematic Analysis

We employed our methods to assess AAA wall displacements and strains in ten patients. For each patient, we used the diastolic and systolic phases of 4D-CTA image data as the moving and fixed images in deformable image registration, respectively.

Table 1 and

Table ***2*** present AAA kinematic analysis results for our ten patients. Table 1 reports, for each patient, the AAA wall geometry, vector plots of wall displacement, contour plots of wall normal displacement, and contour plots of wall strain, using patient-specific contour limits.

Table ***2*** reports, for each patient, the 99th percentile of wall displacement magnitude $U_{\circ}$, 99th percentile of wall normal displacement $u_{\circ}$, and 99th percentile of wall strain $\epsilon_{\circ}$.

At first glance, the strain maps in Table 1 reveal that, in most cases, the highest strain occurs in the anterior regions of AAA wall, with the spine limiting stretching in the posterior region.

Table ***2*** shows that, among all patients, the 99th percentile of actual wall displacement magnitude ranged from 0.70 mm to 1.42 mm, with an average of 1.14 mm and a standard deviation of 0.21 mm, while the 99th percentile of actual wall normal displacement ranged from 0.53 mm to 1.13 mm, with an average of 0.90 mm and a standard deviation of 0.17 mm. The 99th percentile of actual wall strain among all patients ranged from 2.62% to 5.54%, with an average of 4.45% and a standard deviation of 0.87%. The small ratio of standard deviation to average in these results suggest consistency in AAA kinematics across all patients.

To compare strain between the healthy aorta and AAA regions, we computed strain in the healthy aorta regions located proximally and distally to the AAA, for Patient 3 (see Supplementary Material, Section S5, for details). Consistent with previous research (Bell et al., 2014), we observed much higher strain in the healthy aorta than in the AAA, with a 99th percentile strain of 9.21% in the healthy aorta compared to 4.46% in the AAA.

**Table 1.** AAA kinematic analysis results for ten patients. For each patient, the AAA wall geometry, vector plots of wall displacement, contour plots of wall normal displacement, and contour plots of wall strain are displayed. Patient-specific contour limits, including the 99th percentile of wall displacement magnitude $U_\circ$, 99th percentile of wall normal displacement $u_\circ$, and 99th percentile of wall strain $\epsilon_\circ$, are reported for each patient.

| Patient number | AAA wall geometry | Wall displacement vector (mm) $-U_\circ$ 0 $+U_\circ$ 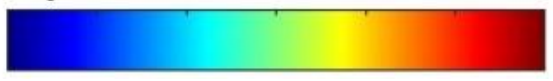 | Wall normal displacement (mm) $-u_\circ$ 0 $+u_\circ$ 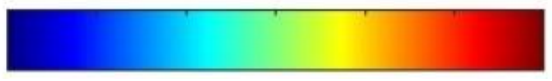 | Wall strain $-\epsilon_\circ$ 0 $+\epsilon_\circ$ 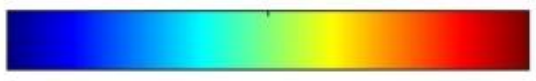 |
|---|---|---|---|---|
| 1 | 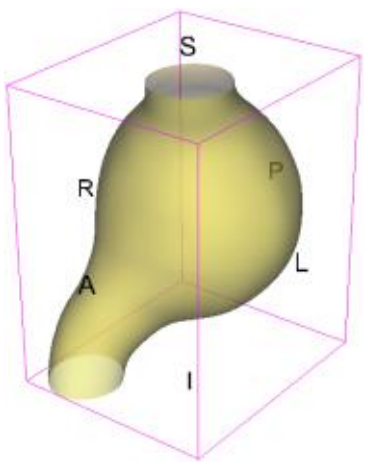  | 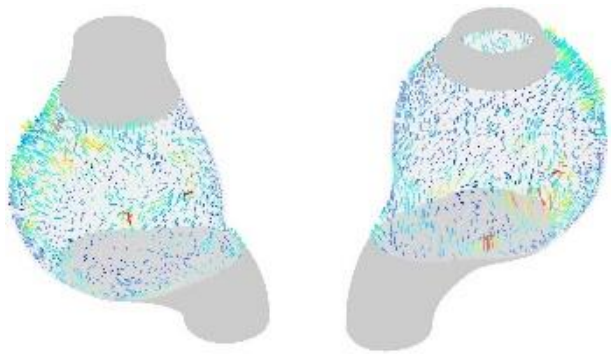 $U_\circ = 1.34$ mm | 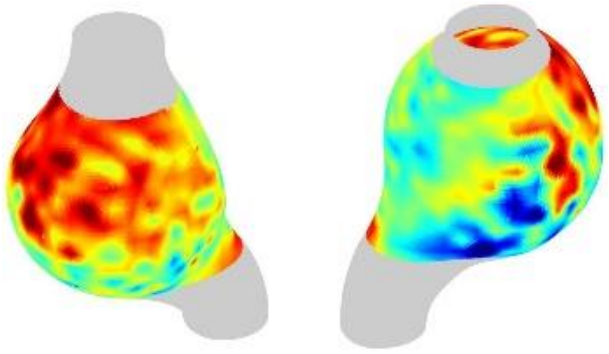 $u_\circ = 0.99$ mm | 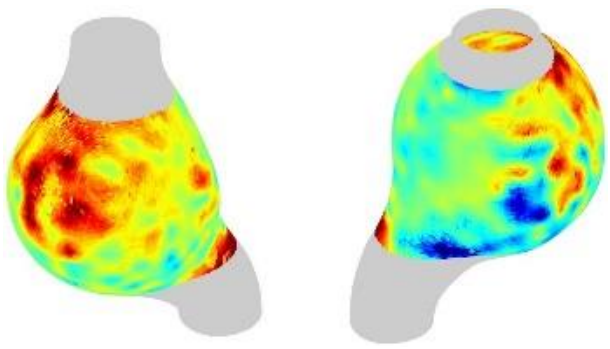 $\epsilon_\circ = 5.54$ % |
| 2 | 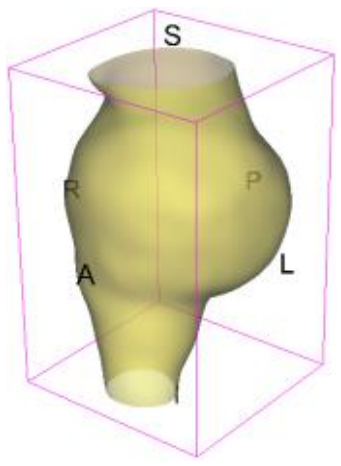  | 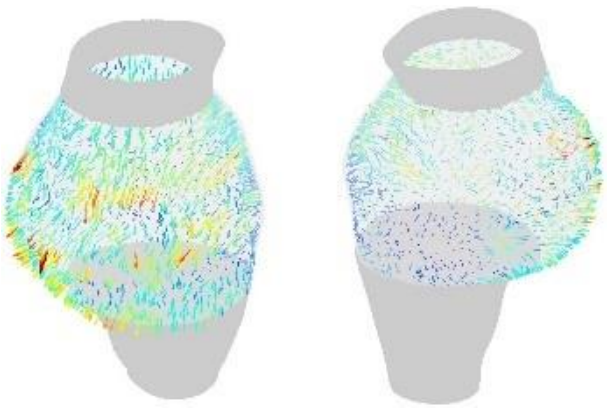 $U_\circ = 1.31$ mm | 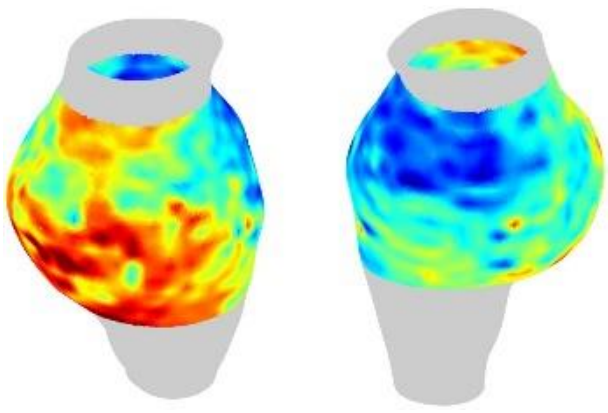 $u_\circ = 1.13$ mm | 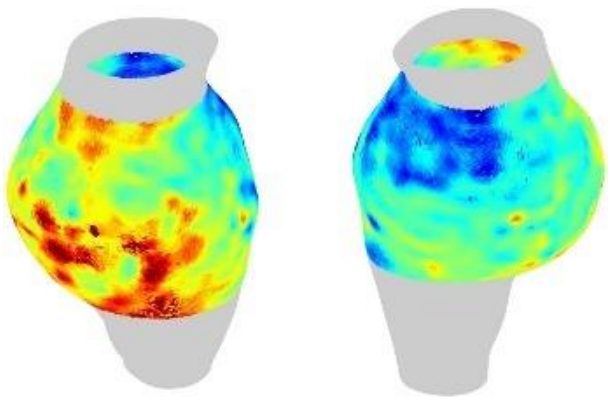 $\epsilon_\circ = 5.23$ % |
| 3 | 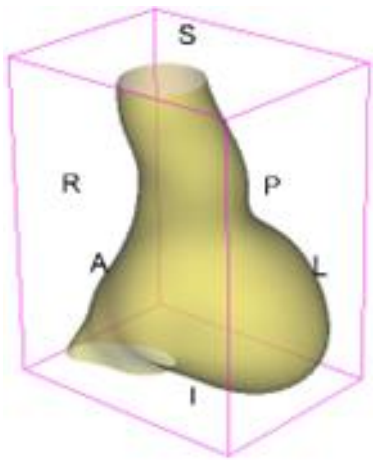  | 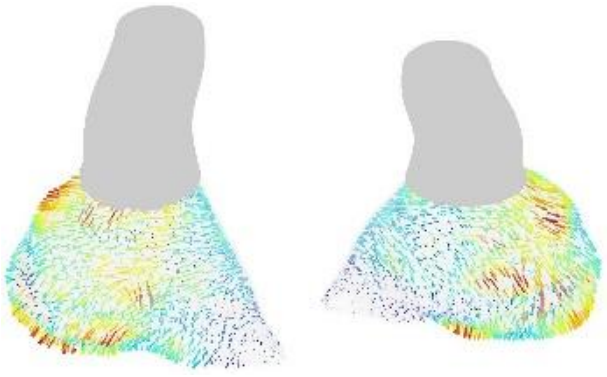 $U_\circ = 1.00$ mm | 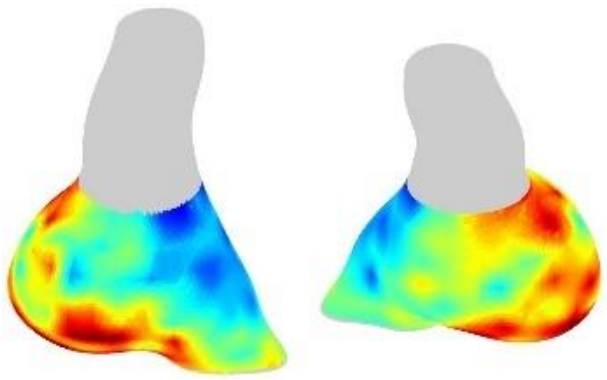 $u_\circ = 0.82$ mm | 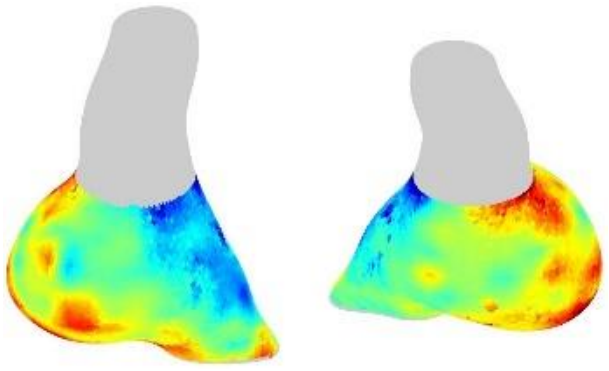 $\epsilon_\circ = 4.46$ % |
| 4 | 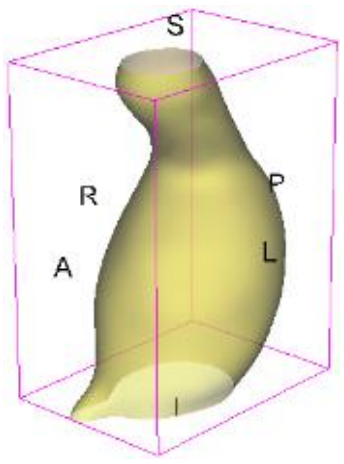  | 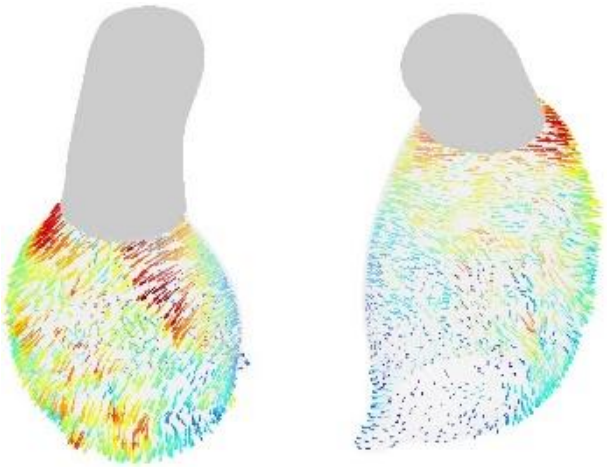 $U_\circ = 1.19$ mm | 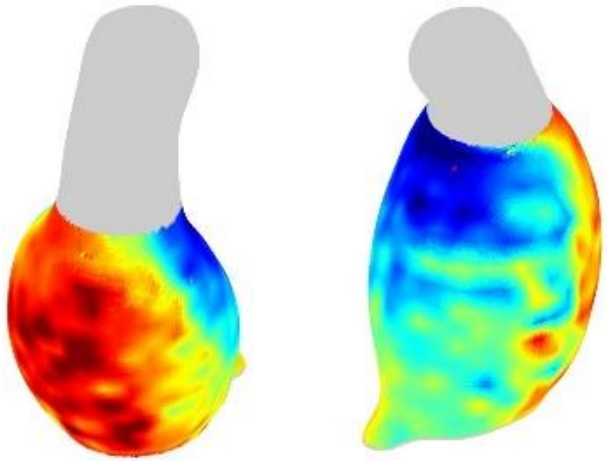 $u_\circ = 0.91$ mm | 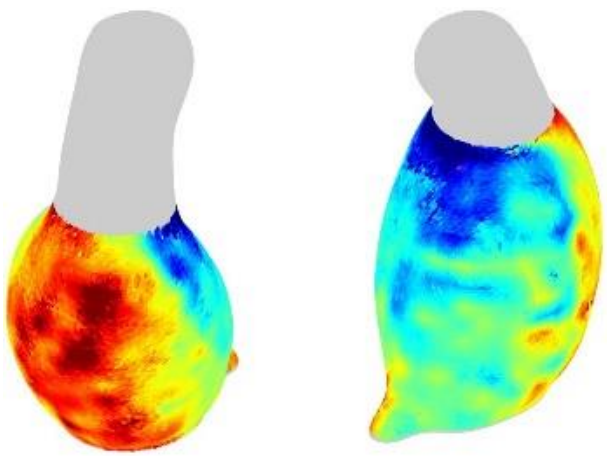 $\epsilon_\circ = 4.01$ % |
| 5 | 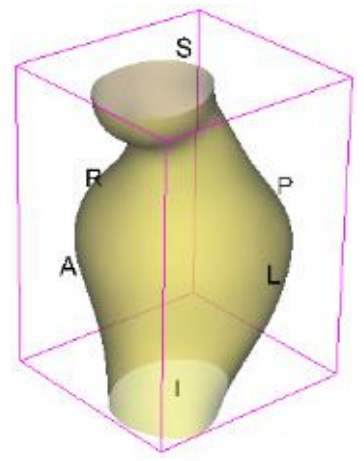 | 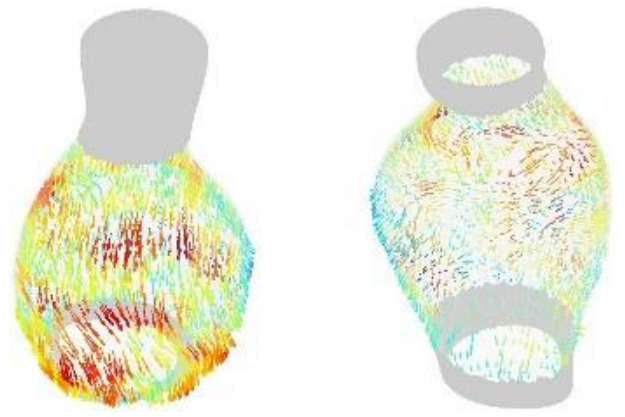 $U_{\circ} = 1.14$ mm | 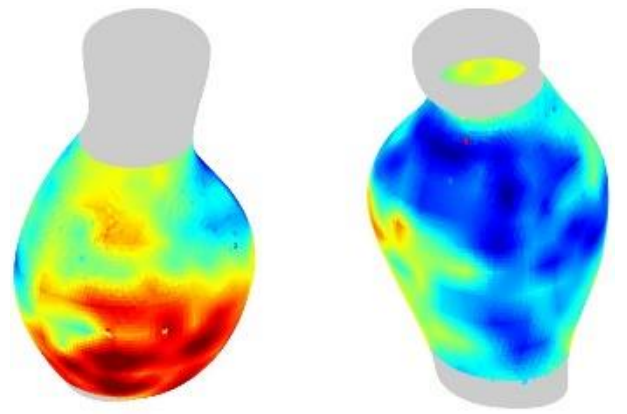 $u_{\circ} = 0.93$ mm | 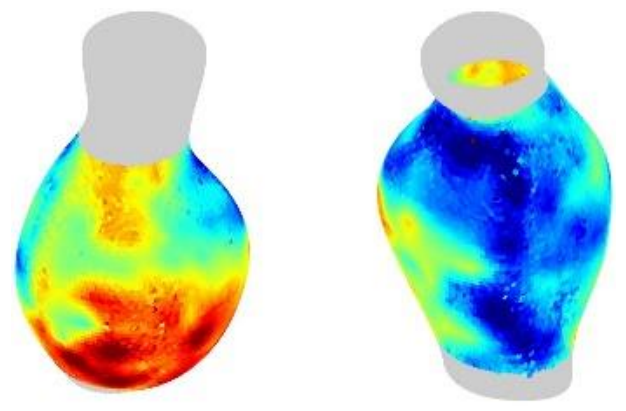 $\epsilon_{\circ} = 3.91$ % |
| 6 | 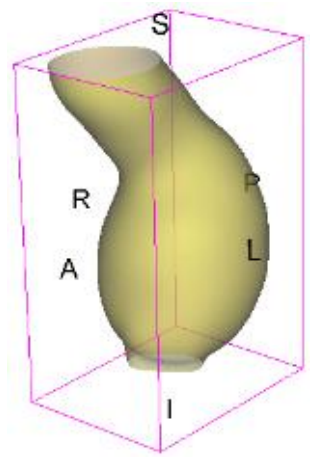 | 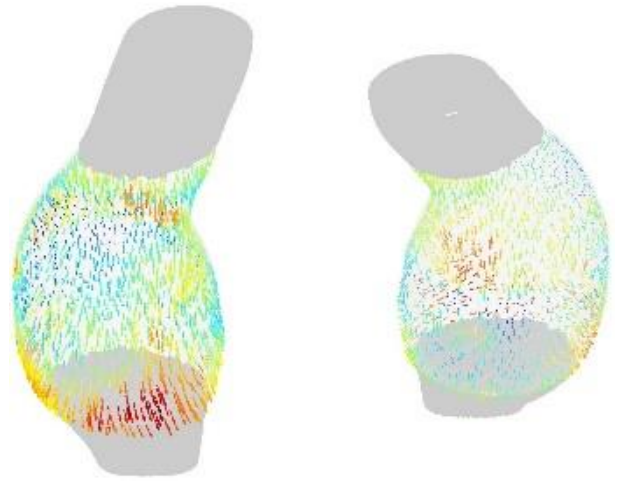 $U_{\circ} = 1.02$ mm | 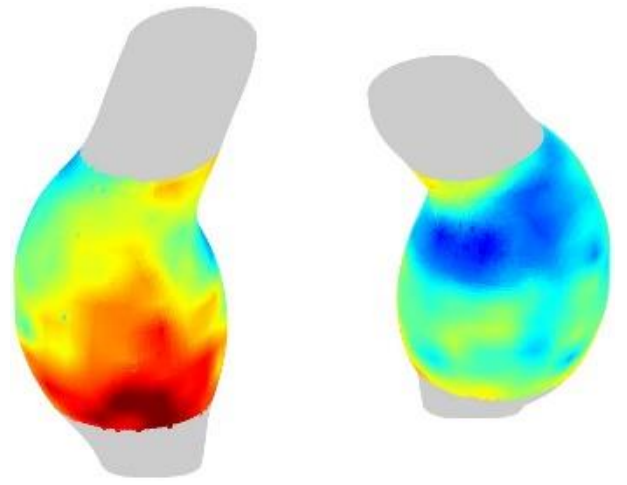 $u_{\circ} = 0.94$ mm | 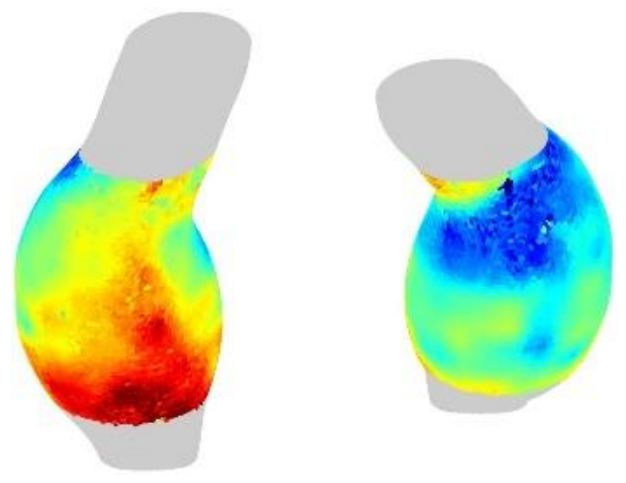 $\epsilon_{\circ} = 4.38$ % |
| 7 | 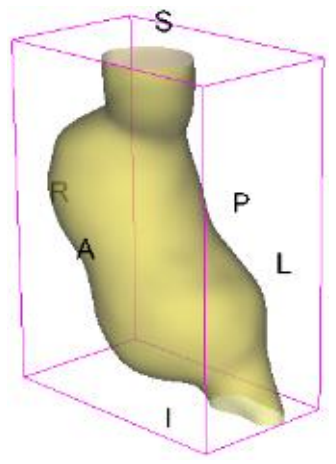 | 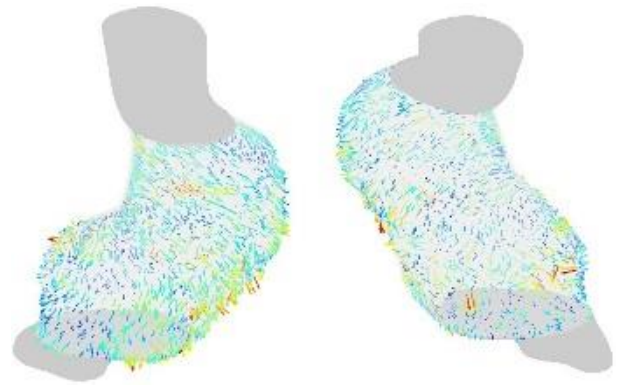 $U_{\circ} = 1.42$ mm | 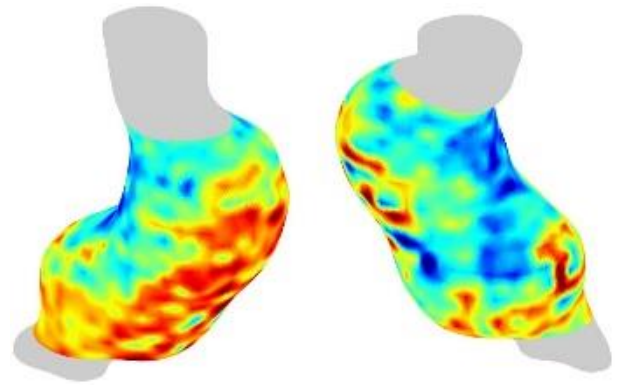 $u_{\circ} = 1.05$ mm | 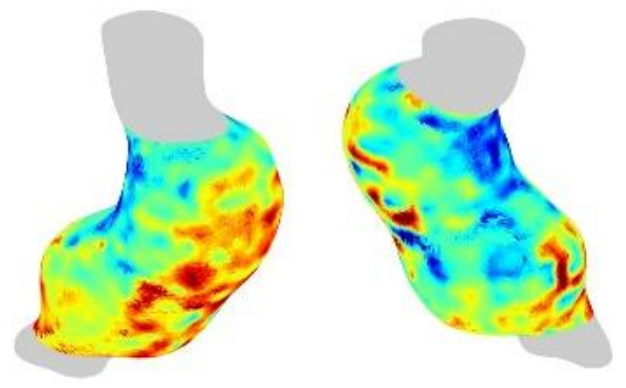 $\epsilon_{\circ} = 5.50$ % |
| 8 | 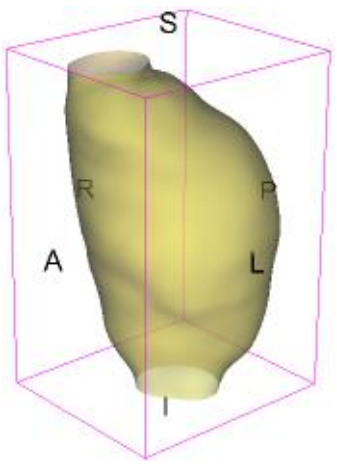 | 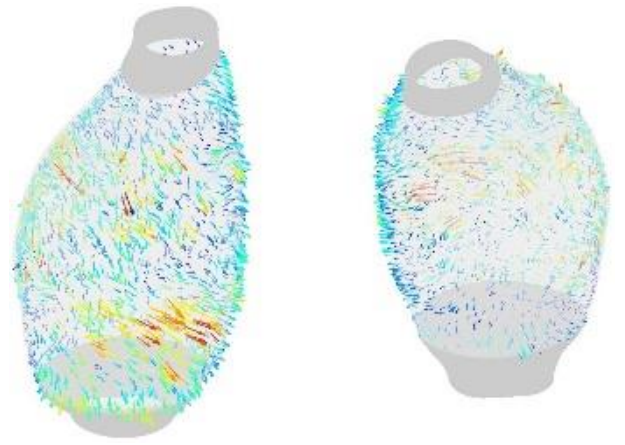 $U_{\circ} = 1.03$ mm | 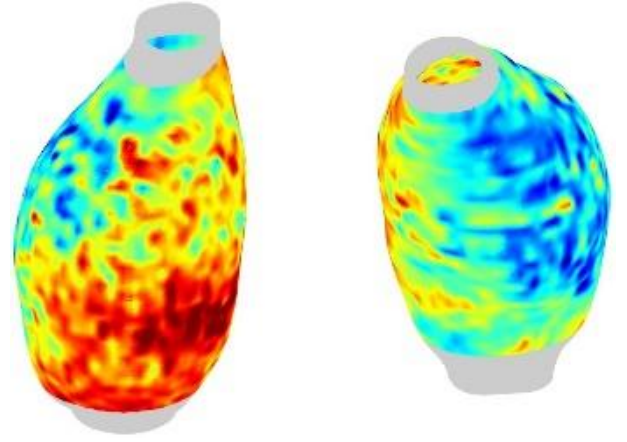 $u_{\circ} = 0.77$ mm | 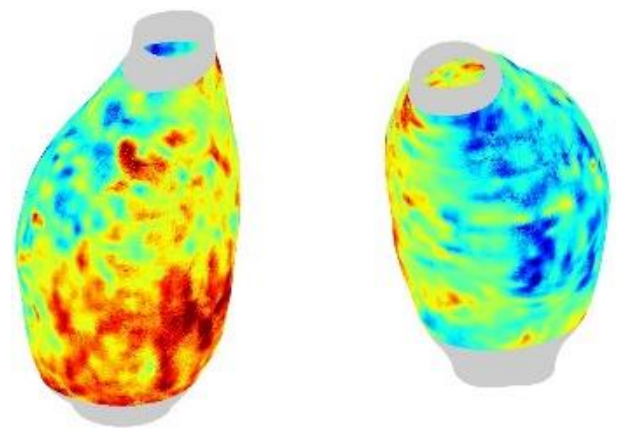 $\epsilon_{\circ} = 4.55$ % |
| 9 | 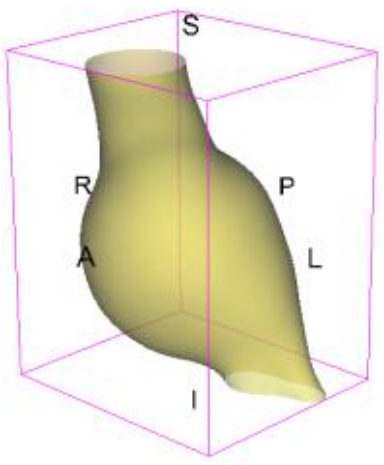 | 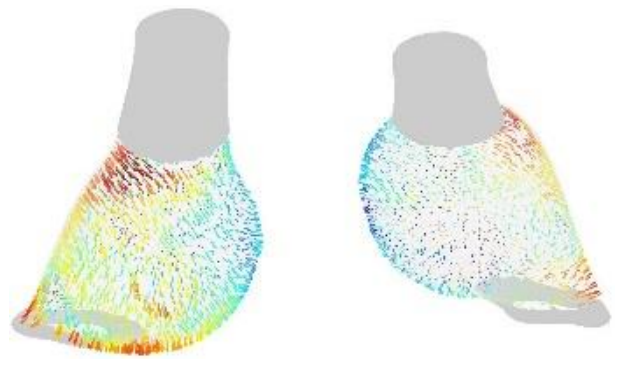 $U_{\circ} = 0.70$ mm | 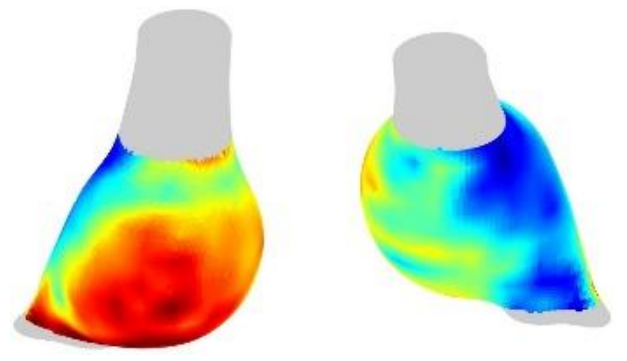 $u_{\circ} = 0.53$ mm | 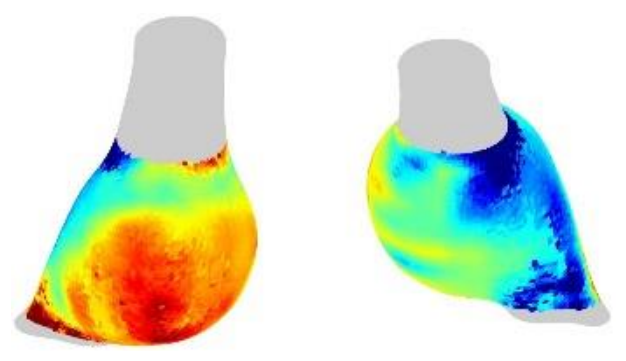 $\epsilon_{\circ} = 2.62$ % |
| 10 | 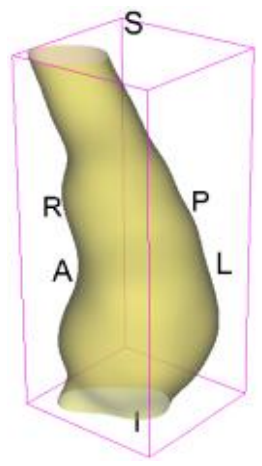 | 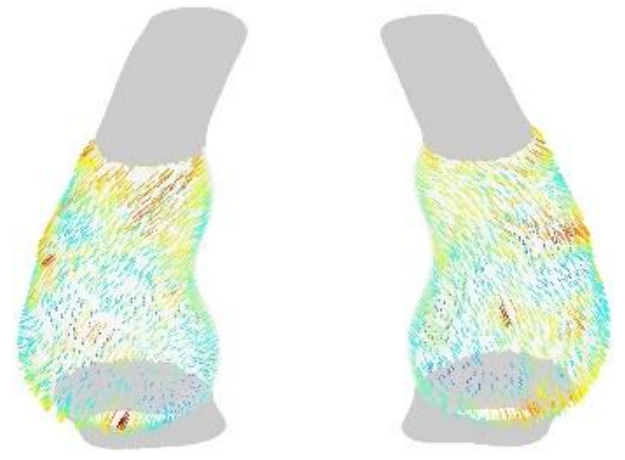 $U_{\circ} = 1.27$ mm | 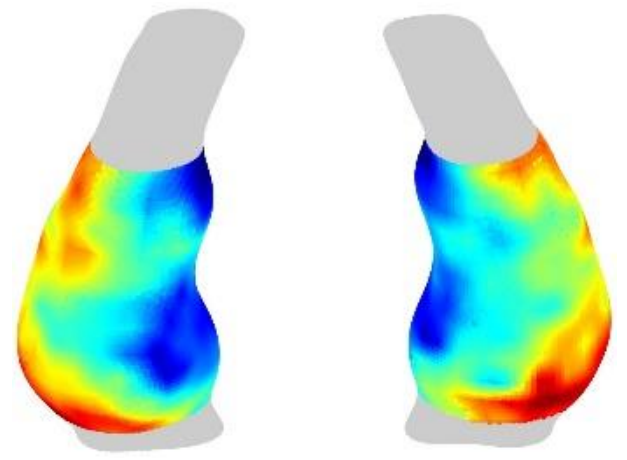 $u_{\circ} = 0.93$ mm | 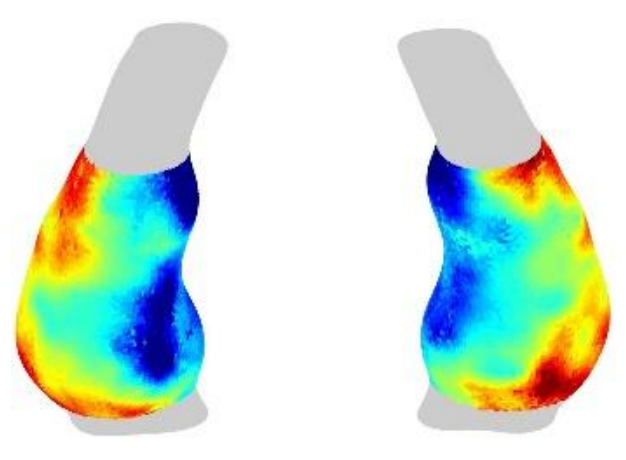 $\epsilon_{\circ} = 4.32$ % |

**Table 2.** AAA kinematic analysis results for ten patients. For each patient, the 99th percentile of wall displacement magnitude $U_\circ$, 99th percentile of wall normal displacement $u_\circ$, and 99th percentile of wall strain $\epsilon_\circ$, are reported.

| Patient number | $\boldsymbol{U_\circ}$ (mm) | $\boldsymbol{u_\circ}$ (mm) | $\boldsymbol{\epsilon_\circ}$ (%) |
|---|---|---|---|
| **1** | 1.34 | 0.99 | 5.54 |
| **2** | 1.31 | 1.13 | 5.23 |
| **3** | 1.00 | 0.82 | 4.46 |
| **4** | 1.19 | 0.91 | 4.01 |
| **5** | 1.14 | 0.93 | 3.91 |
| **6** | 1.02 | 0.94 | 4.38 |
| **7** | 1.42 | 1.05 | 5.50 |
| **8** | 1.03 | 0.77 | 4.55 |
| **9** | 0.70 | 0.53 | 2.62 |
| **10** | 1.27 | 0.93 | 4.32 |
| **Minimum** | 0.70 | 0.53 | 2.62 |
| **Maximum** | 1.42 | 1.13 | 5.54 |
| **Average** | 1.14 | 0.90 | 4.45 |
| **Standard deviation** | 0.21 | 0.17 | 0.87 |

## 5. Conclusions and discussions

We developed an image registration-based approach for patient-specific AAA kinematic analysis, enabling non-invasive, in-vivo measurement of AAA wall displacements and strains from time-resolved 3D computed tomography angiography (4D-CTA) images.

We estimated the wall displacement by registering the diastolic phase image of 4D-CTA image data (moving image), onto the systolic phase image (fixed image). We used deformable image registration with isotropic total variation regularization of displacement. Using planar regression, we established local biological coordinate systems, including local normal and tangential directions to AAA wall, and decomposed the local wall displacement into its normal and tangential components. The component-wise analysis of the wall registration displacement indicated that the tangential displacement component was unreliable, whereas the normal wall displacement was satisfactorily estimated and hence used for strain calculation.

We calculated the local AAA wall strain as the ratio of displacement normal to the wall to its local radius of curvature, with clear interpretation in terms of Green strain components. We estimated the local radius of wall curvature using local cylinder fitting with a Random Sample Consensus (RANSAC) algorithm.

We verified our approach by demonstrating an excellent match between the registration and biomechanically established ground truth wall displacements and strains.

We assessed wall displacements and strain in ten patients. Among all patients, the 99th percentile of circumferential wall strain ranged from 2.62% to 5.54%, with an average of 4.45% and a standard deviation of 0.87%. Noteworthy was that the spine restricted AAA stretching in the posterior region. Furthermore, we observed that AAA wall strains were significantly lower than those of a healthy aorta.

A notable strength of our kinematic and purely image-based approach is that various biomechanical uncertainties related to e.g. aortic calcification, the extent of intraluminal thrombus, and patient-specific wall properties have no impact on the accuracy of our results as their effects are embedded in the images. We believe that rapid progress in patient-specific applications of biomechanics can only be achieved when analysis results are not sensitive to the uncertain mechanical properties of tissues.

Nevertheless, limitations of our method need to be clearly articulated. In current clinical practice of AAA diagnosis and treatment, 4D-CTA is rarely used. However, it is a common imaging modality in cardiology and therefore easily translatable to vascular disease management.

As image gradients in direction tangential to AAA wall are small, our image registration method is unable to capture reliably the tangential components of AAA wall displacements. Therefore, in strain estimation we must rely only on the normal component of the displacement vector and not on all three displacement vector components. This clearly inhibits the ability to identify a full finite strain tensor.

Despite these limitations, our approach offers the clear advantage of evaluating aneurysm displacements and strains locally, allowing for point-by-point evaluation.

## Acknowledgements

This work was supported by the Australian National Health and Medical Research Council NHMRC Ideas grant No. APP2001689. Contributions of Christopher Wood and Jane Polce, radiology technicians at the Medical Imaging Department, Fiona Stanley Hospital (Murdoch, Western Australia) to patient image acquisition at Fiona Stanley Hospital, and Dr Farah Alkhatib's (The University of Western Australia) assistance in patient recruitment and acquisition of these images are gratefully acknowledged.

## List of figure legends

| |
|---|
| **Figure 1.** Patient 1's AAA: (a) Cropped 3D-CTA image in the diastolic phase, (b) AAA model, and (c) Hexahedral finite element mesh of AAA model. |
| **Figure 2.** Smooth field of unit normal vectors on AAA surface using planar regression |
| **Figure 3.** Contour plots for AAA wall radius of curvature and vector plots for orientation of the cylinder fitted to a local neighbourhood of points. |
| **Figure 4.** Contour plots of displacement magnitude and vector plots of displacement from the biomechanics-calculated AAA wall deformation used as ground truth for method verification. |
| **Figure 5.** Overlay of the Canny edges for (a) moving (diastolic or undeformed AAA) image versus fixed (synthetic systolic or deformed AAA) image, and (b) registered image versus fixed image, on a selected R-A plane. Fixed image Canny edges are in green, moving image and registered image Canny edges are in magenta, and white shows overlap. |
| **Figure 6.** Contour plots of AAA wall displacement magnitude, normal displacement, and tangential displacement for (a) registration displacement field and (b) ground truth displacement field. (c) Contour plot of the difference between registration and ground truth displacements for displacement magnitude, normal displacement, and tangential displacement. |
| **Figure 7.** Scatter plots of registration versus ground truth displacements for (a) displacement magnitude, (b) normal displacement, and (c) tangential displacement. Each scatter plot is coloured by the angle between the corresponding registration and ground truth displacement vectors. The R-squared value and Normalized Root Mean Square Error (NRMSE) of the identity line fit (y = x) are shown within each plot. In each case, the null hypothesis that the difference is not zero is confidentially rejected with p-values less than $10^{-15}$. |
| **Figure 8.** Contour plots of (a) wall strain from registration, (b) ground truth wall strain, and (c) the difference between strain from registration and ground truth strain. (d) Scatter plot of wall strain from registration versus ground truth wall strain, with the R-squared value and Normalized Root Mean Square Error (NRMSE) of the identity line fit (y = x) shown within the plot. The null hypothesis that the difference is not zero is confidentially rejected with p-values less than $10^{-15}$. |

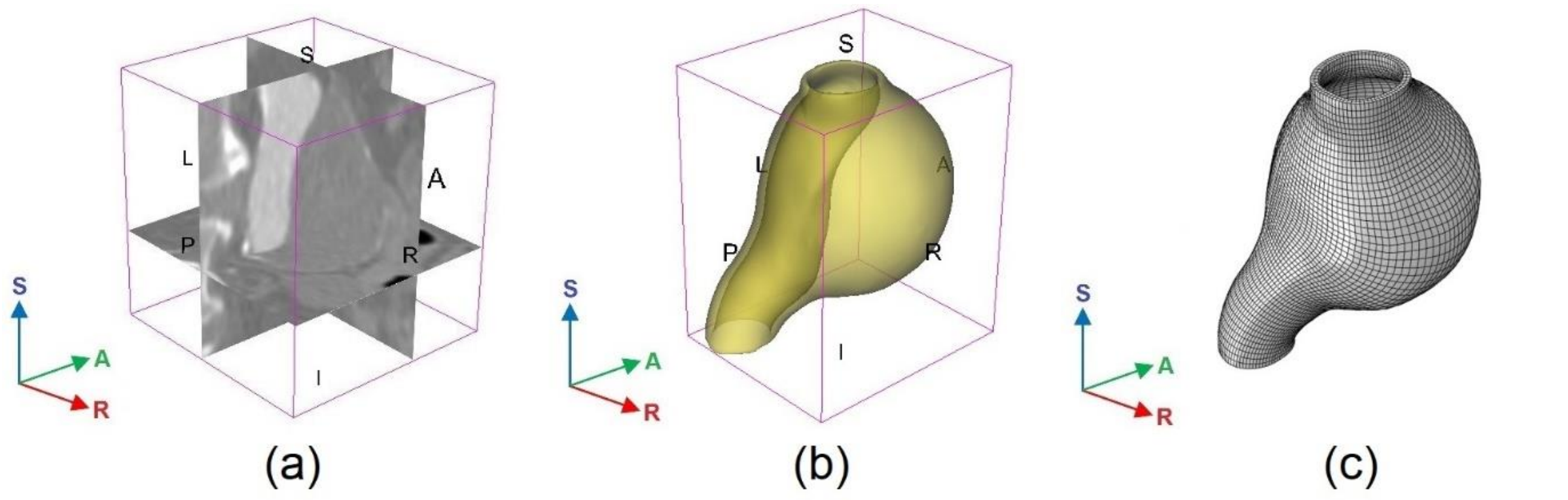

**Figure 1.** Patient 1's AAA: (a) Cropped 3D CTA image in the diastolic phase, (b) AAA model, and (c) Hexahedral finite element mesh of AAA model.

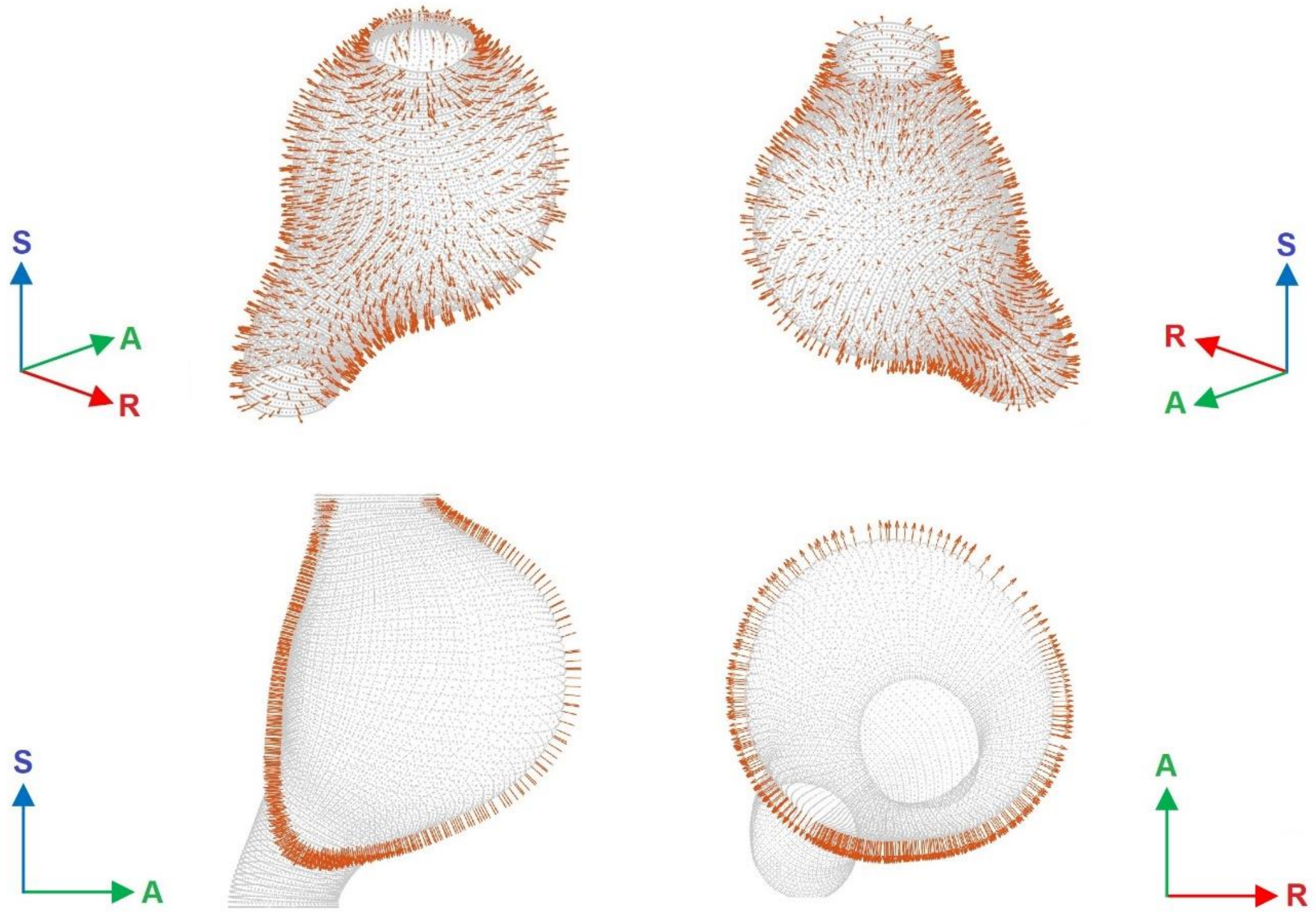

**Figure 2.** Smooth field of unit normal vectors on AAA surface using planar regression

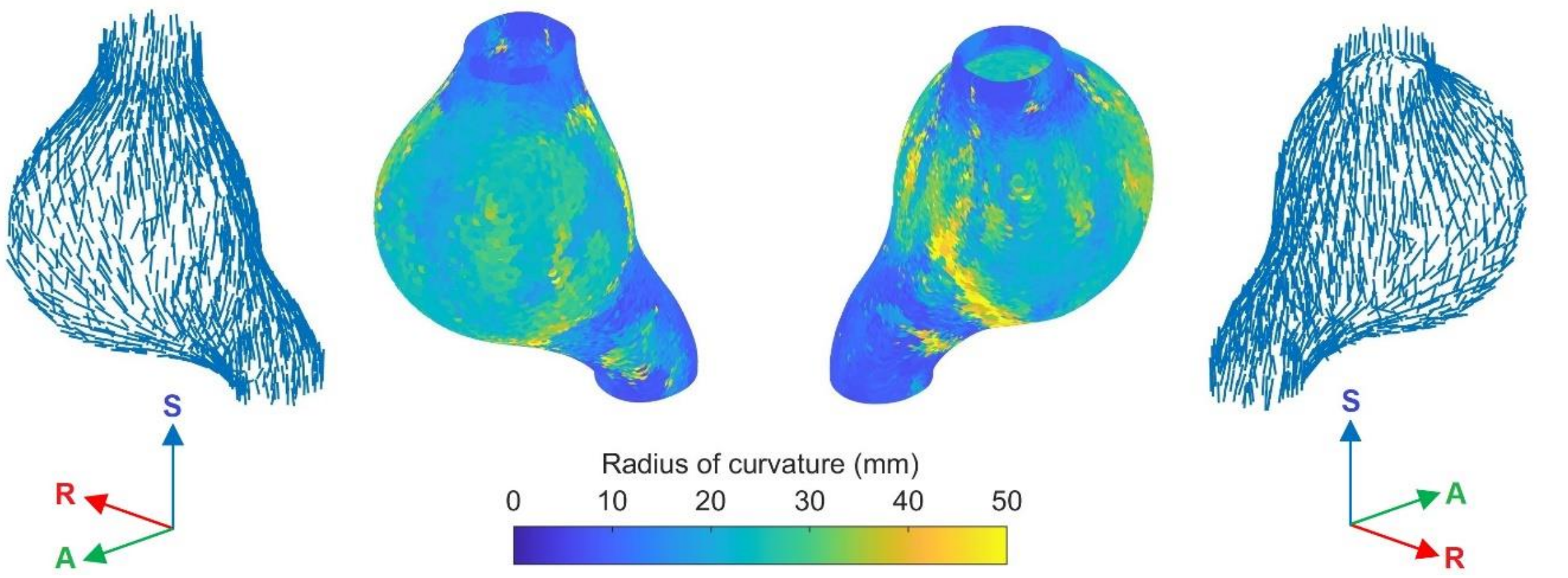

**Figure 3.** Contour plots for AAA wall radius of curvature and vector plots for orientation of the cylinder fitted to a local neighbourhood of points.

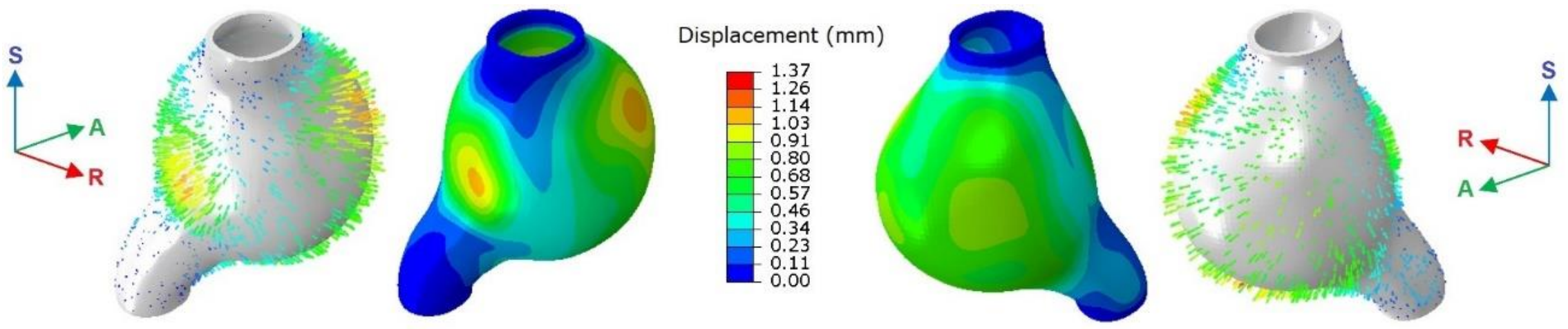

**Figure 4.** Contour plots of displacement magnitude and vector plots of displacement from the biomechanics-calculated AAA wall deformation used as ground truth for method verification.

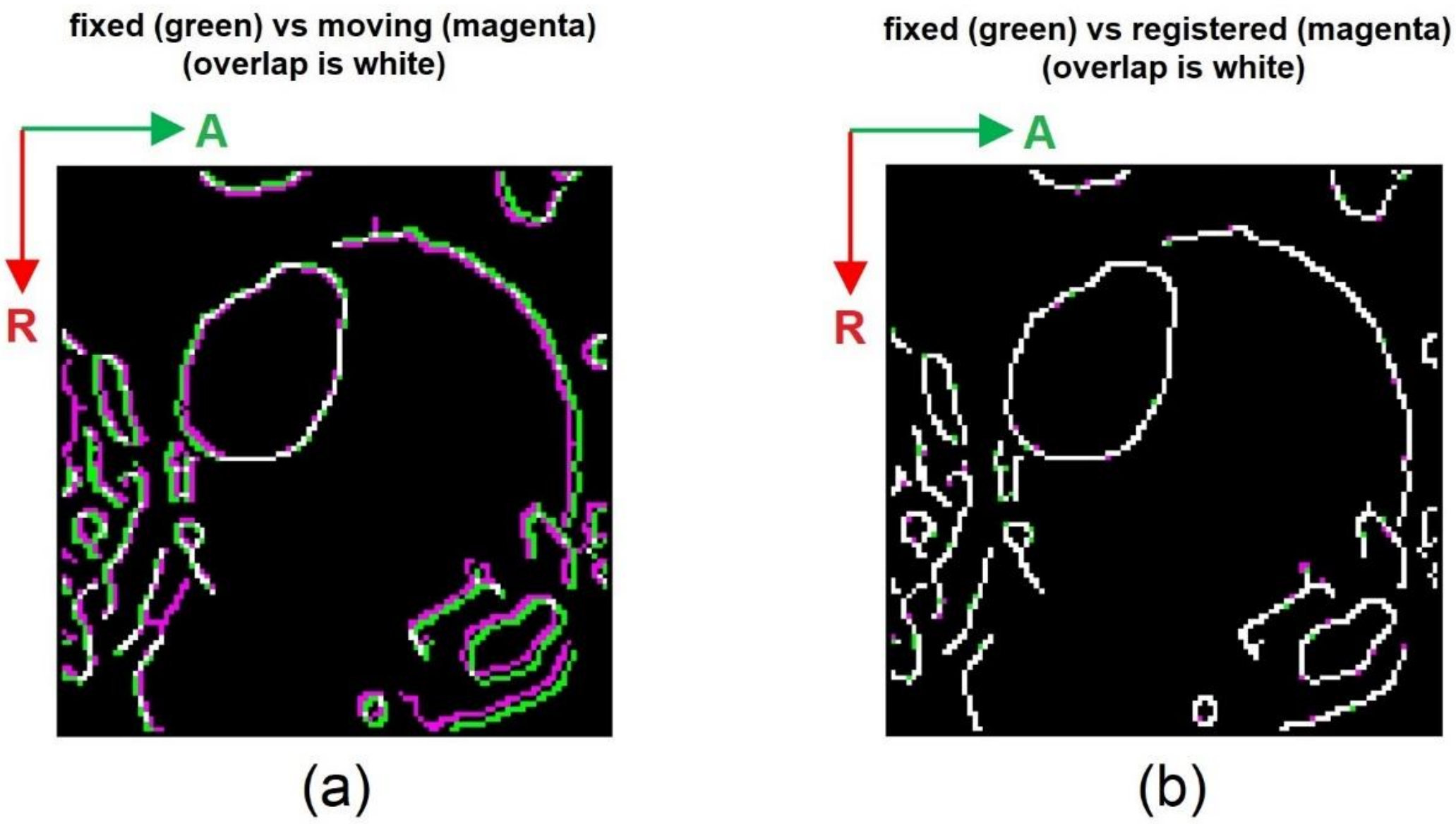

**Figure 5.** Overlay of the Canny edges for (a) moving (diastolic or undeformed AAA) image versus fixed (synthetic systolic or deformed AAA) image, and (b) registered image versus fixed image, on a selected R-A plane. Fixed image Canny edges are in green, moving image and registered image Canny edges are in magenta, and white shows overlap.

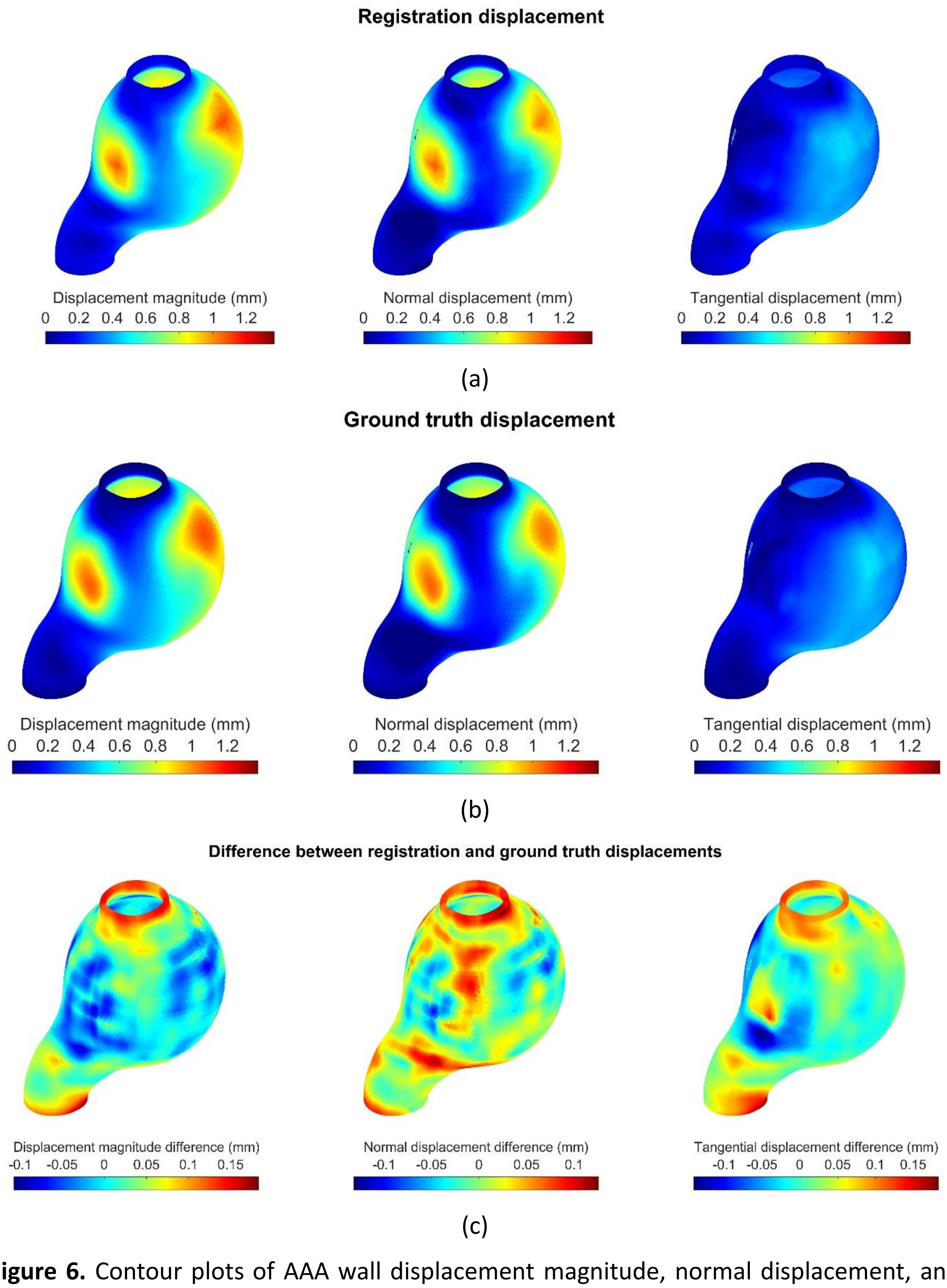

**Figure 6.** Contour plots of AAA wall displacement magnitude, normal displacement, and tangential displacement for (a) registration displacement field and (b) ground truth displacement field. (c) Contour plot of the difference between registration and ground truth displacements for displacement magnitude, normal displacement, and tangential displacement.

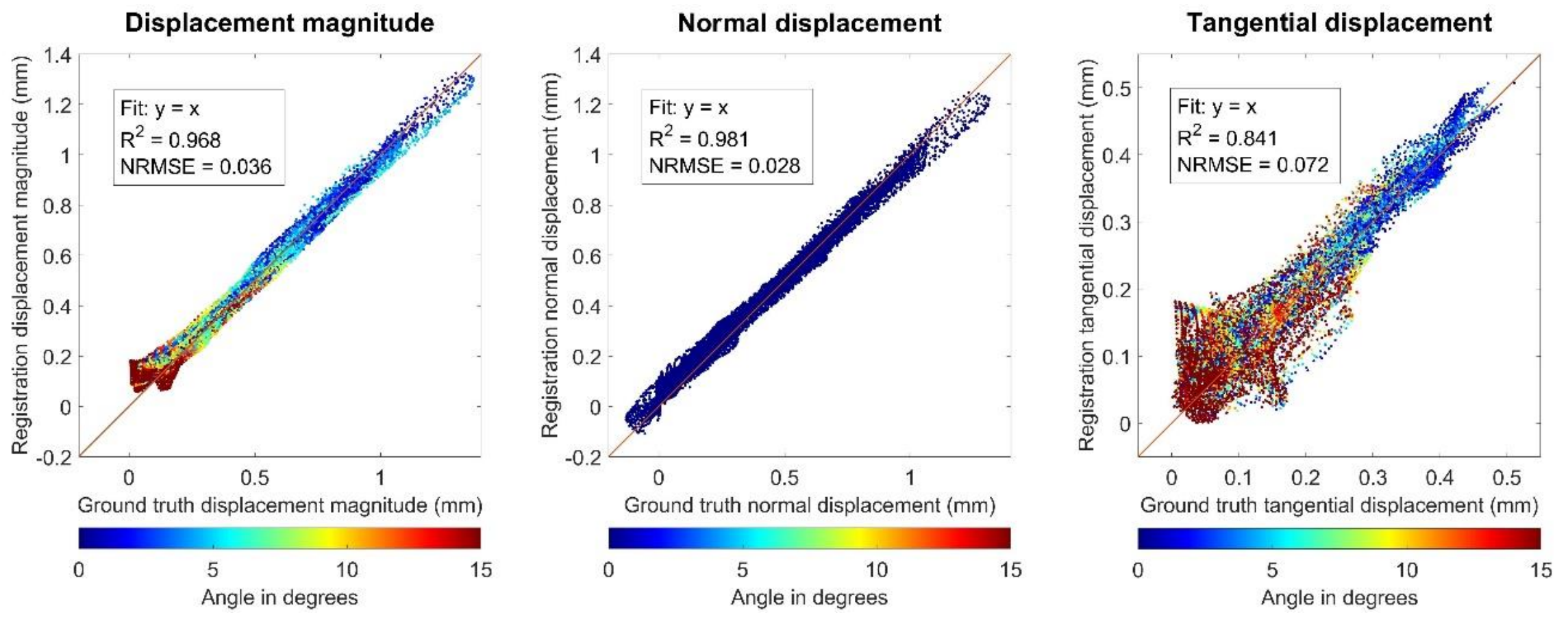

**Figure 7.** Scatter plots of registration versus ground truth displacements for (a) displacement magnitude, (b) normal displacement, and (c) tangential displacement. Each scatter plot is coloured by the angle between the corresponding registration and ground truth displacement vectors. The R-squared value and Normalized Root Mean Square Error (NRMSE) of the identity line fit (y = x) are shown within each plot. In each case, the null hypothesis that the difference is not zero is confidentially rejected with p-values less than $10^{-15}$.

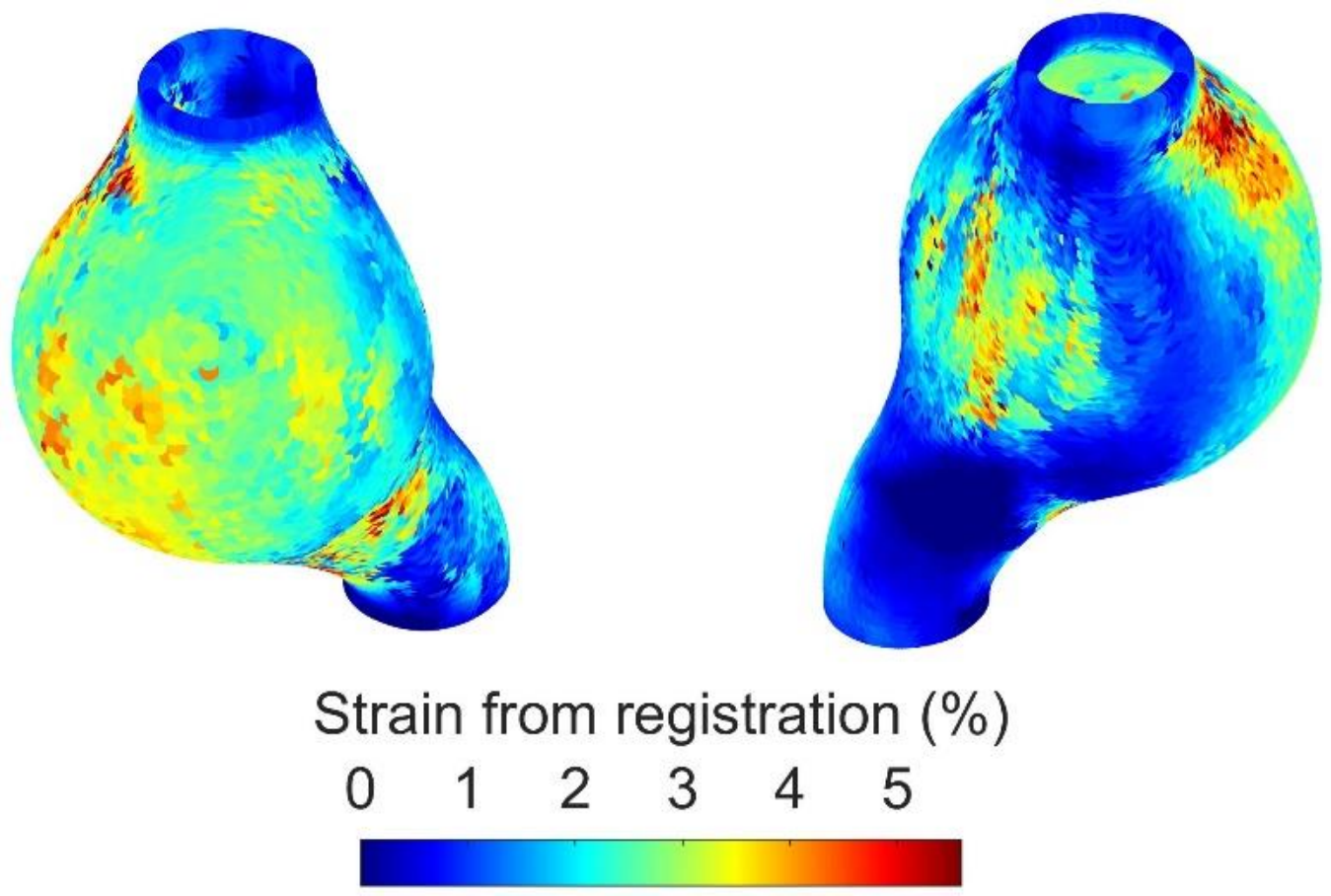
Strain from registration (%)
0 1 2 3 4 5

(a)

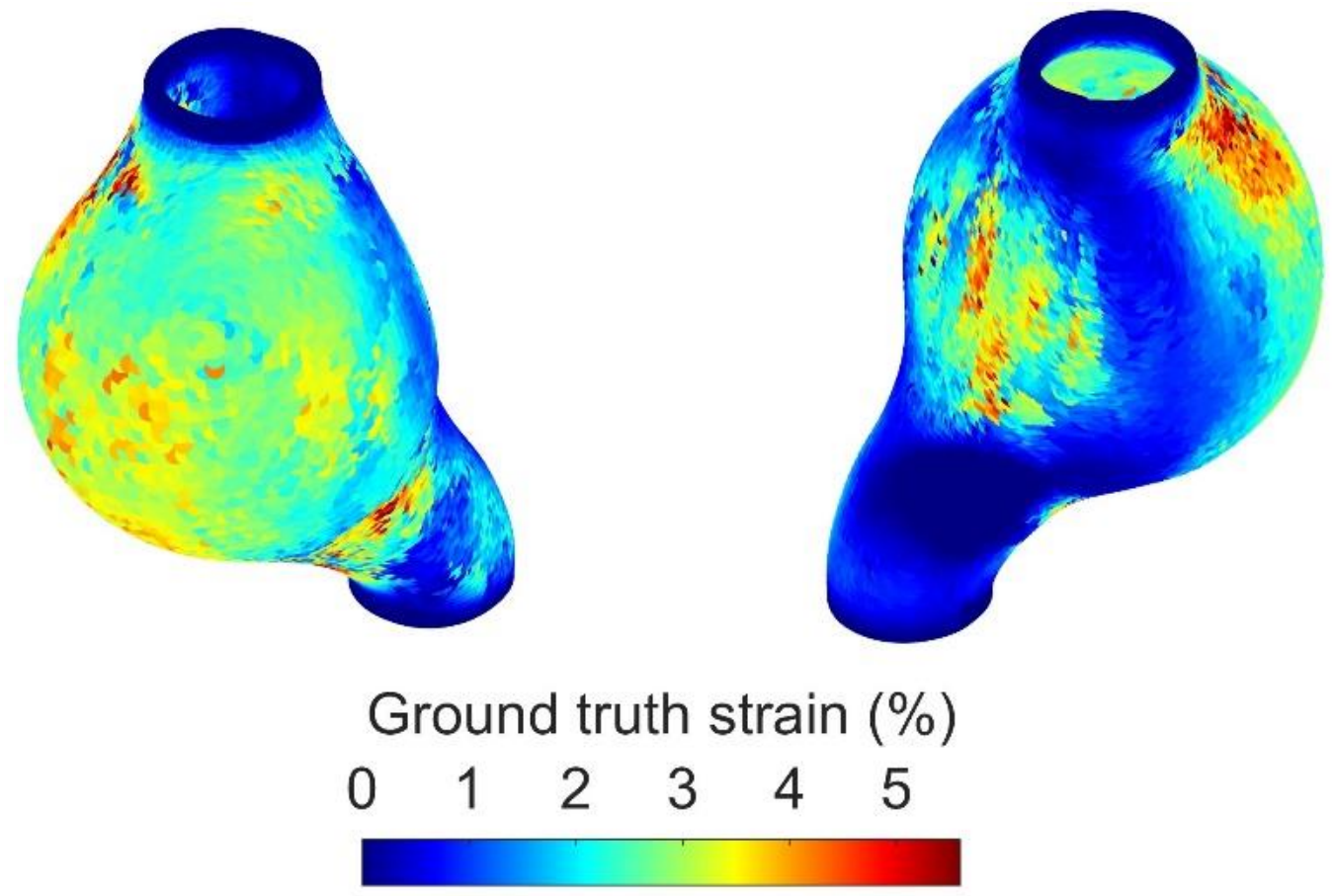
Ground truth strain (%)
0 1 2 3 4 5

(b)

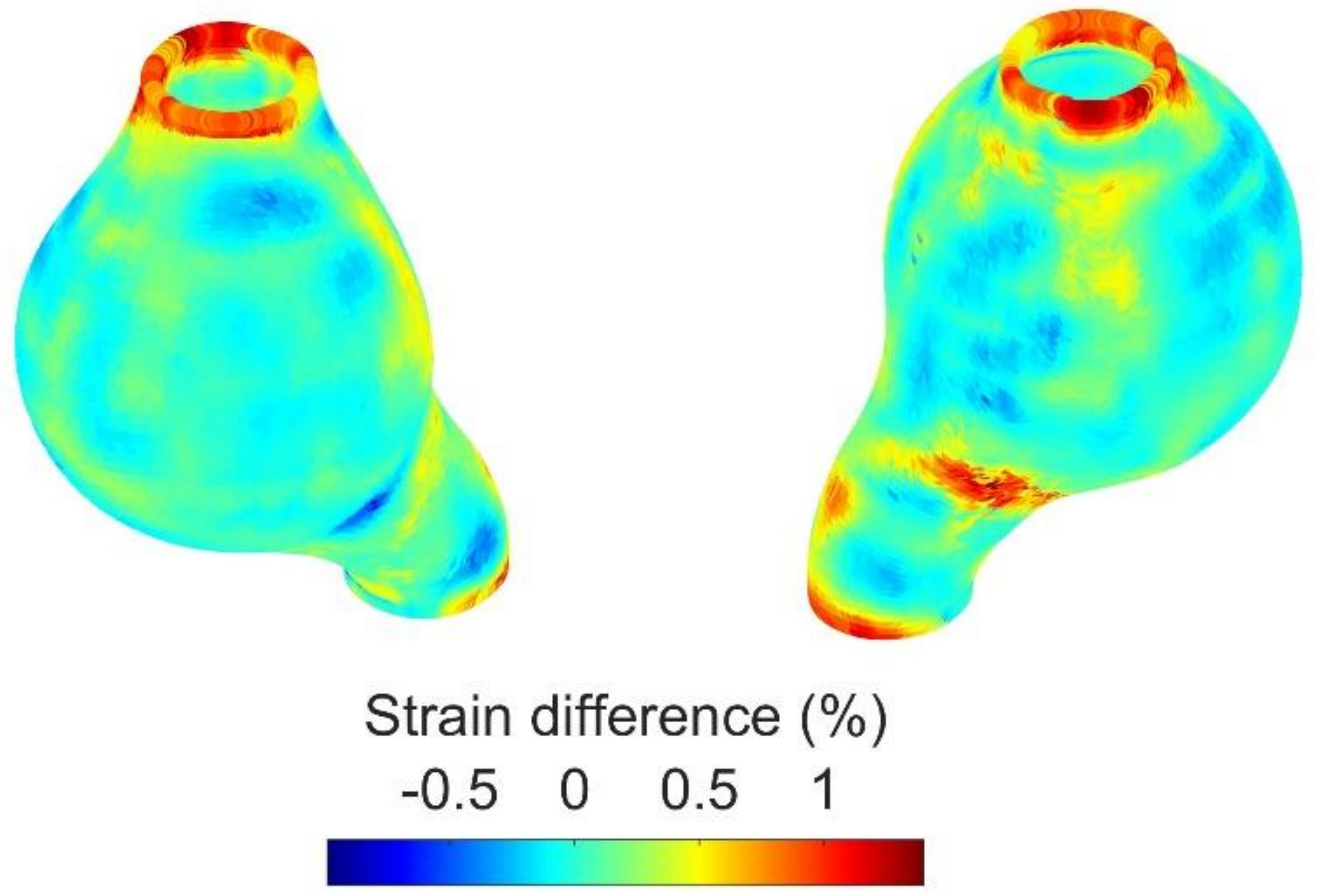
Strain difference (%)
-0.5 0 0.5 1

(c)

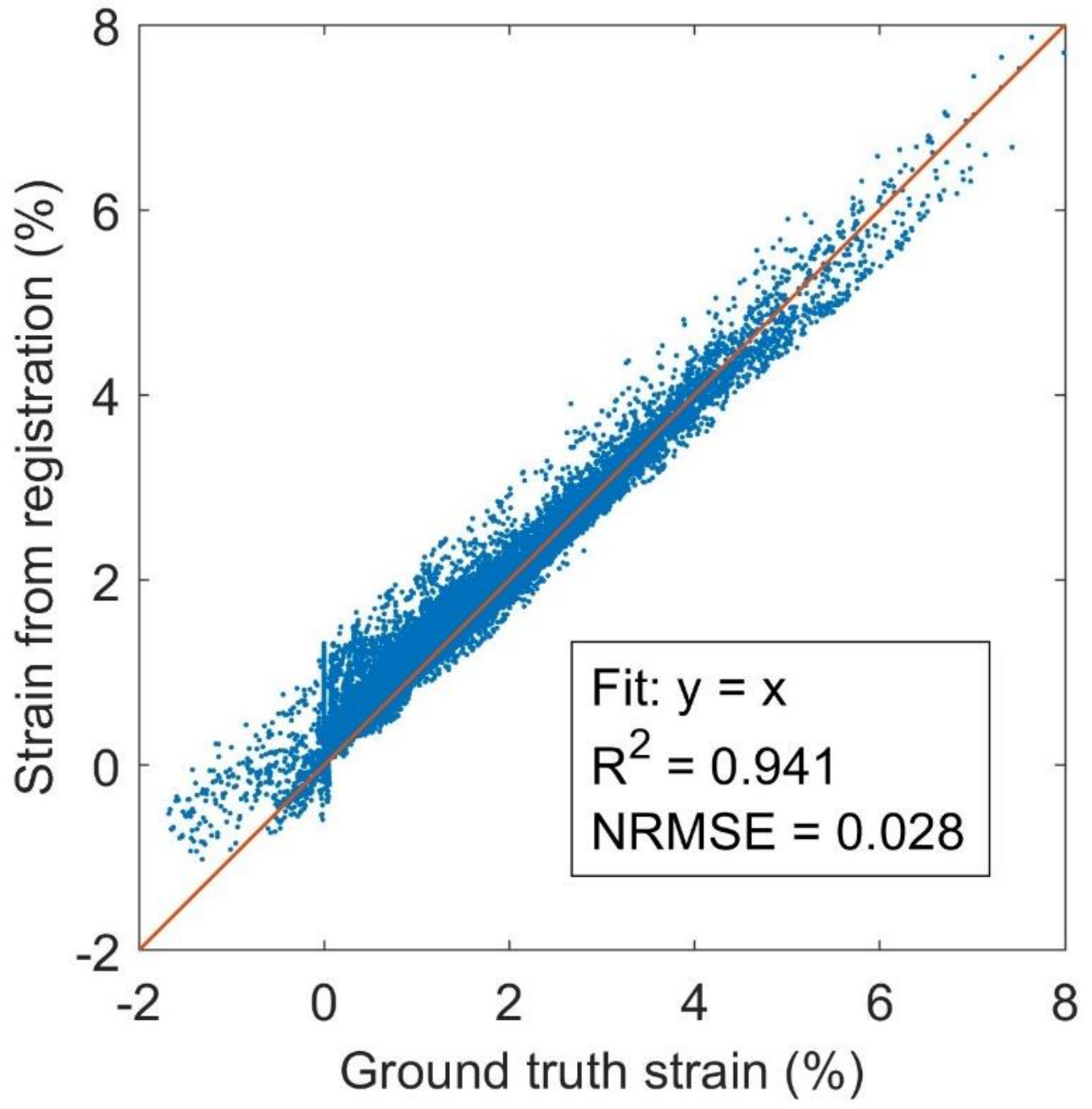

(d)

**Figure 8.** Contour plots of (a) wall strain from registration, (b) ground truth wall strain, and (c) the difference between strain from registration and ground truth strain. (d) Scatter plot of wall strain from registration versus ground truth wall strain, with the R-squared value and Normalized Root Mean Square Error (NRMSE) of the identity line fit (y = x) shown within the plot. The null hypothesis that the difference is not zero is confidentially rejected with p-values less than $10^{-15}$.

# Supplementary Material

## S1. Image data

We used anonymized contrast-enhanced 4D-CTA image datasets, each comprising ten 3D volume frames per cardiac cycle, from ten patients diagnosed with AAA. The patients were recruited at Fiona Stanley Hospital in Western Australia, and informed consent was obtained prior to their participation. The study was conducted in accordance with the Declaration of Helsinki, and the protocol was approved by Human Research Ethics and Governance at South Metropolitan Health Service (HREC-SMHS) (approval code RGS3501), and by Human Research Ethics Office at The University of Western Australia (approval code RA/4/20/5913).

Table S1 provides the image dimensions and resolutions for each patient, along with the maximum AAA diameter, which currently serves as the primary criterion for surgical intervention. The image size and resolution varied among individuals depending on the permitted safe radiation dose, which was determined by the patient's weight and height. The average age of the patients was 77 years. Of the patients, 80% were men and 20% were women, consistent with the higher prevalence of AAA in men compared to women.

**Table S1.** Computed tomography angiography (CTA) image dimensions and resolutions for patients with abdominal aortic aneurysm (AAA), along with the maximum AAA diameter for each patient.

| Patient No. | Image dimensions (pixels) | Image resolution (mm) | AAA maximum diameter (mm) |
|---|---|---|---|
| 1 | 512 × 512 × 143 | 0.63 × 0.63 × 1.00 | 50.9 |
| 2 | 512 × 512 × 169 | 0.63 × 0.63 × 1.50 | 67.2 |
| 3 | 256 × 256 × 254 | 1.18 × 1.18 × 1.00 | 65.8 |
| 4 | 256 × 256 × 177 | 0.81 × 0.81 × 1.50 | 59.3 |
| 5 | 256 × 256 × 482 | 1.82 × 1.82 × 1.00 | 52.7 |
| 6 | 256 × 256 × 488 | 1.53 × 1.53 × 1.00 | 47.1 |
| 7 | 512 × 512 × 160 | 0.63 × 0.63 × 1.00 | 57.3 |
| 8 | 512 × 512 × 155 | 0.31 × 0.31 × 1.00 | 44.2 |
| 9 | 256 × 256 × 170 | 1.25 × 1.25 × 1.00 | 50.2 |
| 10 | 256 × 256 × 184 | 1.25 × 1.25 × 1.00 | 52.3 |

## S2. Regularized image registration algorithm

Figure S1 shows the flowchart and basic components of the regularized image registration algorithm.

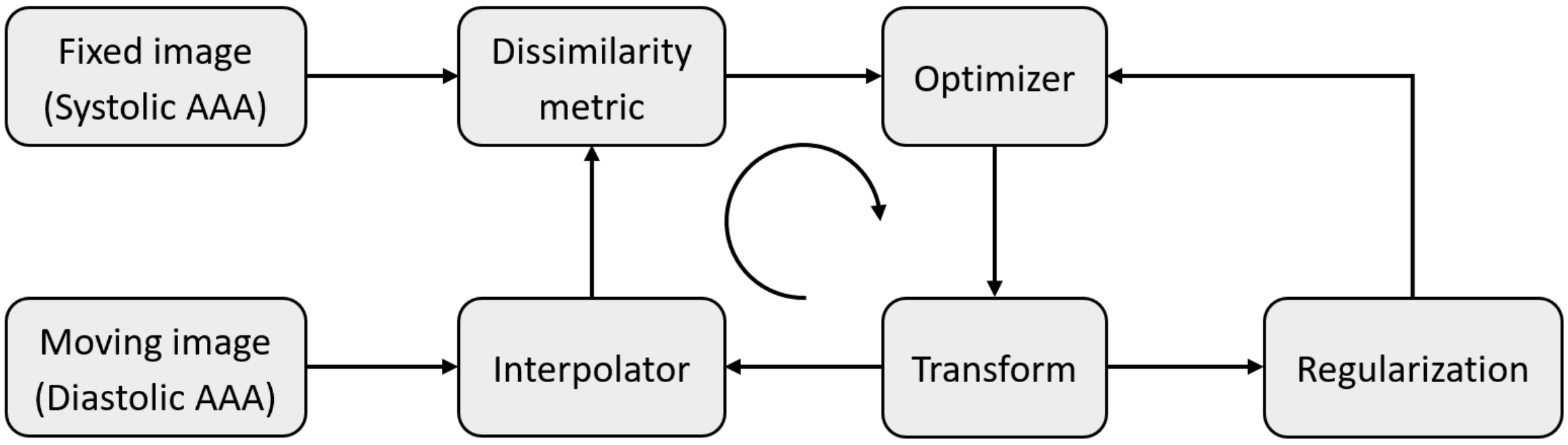

**Figure S1.** Flowchart and basic components of the regularized image registration algorithm.

## S3. Surface normal estimation via 3D plane fitting

In first-order 3D plane fitting (Berkmann and Caelli, 1994), finding the surface normal at a query point is simplified by estimating the normal of a tangent plane through a least-square plane fitting approach (Shakarji, 1998). The solution for estimating the surface normal is therefore reduced to the Principal Component Analysis (PCA) of a covariance matrix created from the nearest neighbours of the query point. Once the neighbouring points set $\mathcal{P}^k$ of a given query point $\boldsymbol{p}$ in the point cloud are identified, the covariance matrix $\boldsymbol{C}$ for the query point $\boldsymbol{p}$ is assembled as

$$\boldsymbol{C} = \frac{1}{k}\sum_{i=1}^{k}(\boldsymbol{p}_i - \overline{\boldsymbol{p}}) \otimes (\boldsymbol{p}_i - \overline{\boldsymbol{p}}) \tag{S1}$$

where $k$ is the number of neighboring points $\boldsymbol{p}_i \in \mathcal{P}^k$ in the vicinity of $\boldsymbol{p}$ and $\overline{\boldsymbol{p}}$ represents the centroid of these neighboring points. Matrix $\boldsymbol{C}$ is symmetric and positive semi-definite with real eigenvalues $\lambda_j$ for $j = 1, 2, 3$. The corresponding eigenvectors $\boldsymbol{v}_j$ form an orthogonal basis, corresponding to the principal components of $\mathcal{P}^k$. If $0 \leq \lambda_1 \leq \lambda_2 \leq \lambda_3$, the eigenvector $\boldsymbol{v}_1$ corresponding to the smallest eigenvalue $\lambda_1$ is the normal.

## S4. Normality assessment

We overlaid the histogram distribution of the point-by-point difference between the normal displacement from registration and the ground truth normal displacement with the histogram of the normal distribution fit, as shown in Figure S2a. This figure suggests that the displacement differences closely follow a normal distribution centred around zero.

We used quantile-quantile (Q-Q) plot introduced by Wilk and Gnanadesikan (1968), to evaluate whether the normal displacement from registration and the ground truth normal displacement have similar distributions, as shown in Figure S2b. Each point in this Q-Q plot represents a quantile of the registration displacement plotted against the corresponding quantile of the ground truth displacement. The closer the points are to the 45-degree red identity line, the more similar the distributions (Wilk and Gnanadesikan, 1968). In Figure S2b, the points largely follow the identity line, suggesting a good agreement between the registration and ground truth displacement distributions.

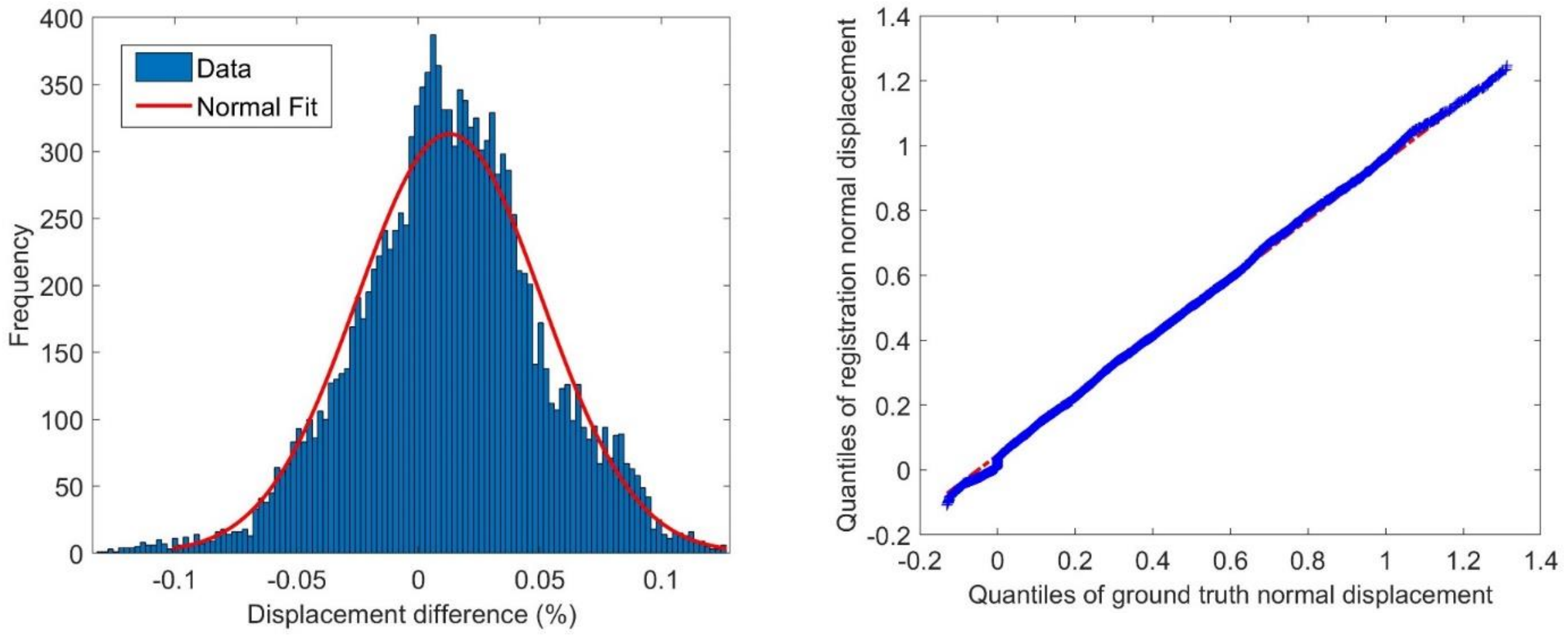

(a) (b)

**Figure S2.** Normality assessment of displacement difference between registration and ground truth: (a) Distribution of point-by-point difference between registration normal displacement and ground truth normal displacement, (b) QQ plot of registration normal displacement versus ground truth normal displacement.

## S5. Strain analysis of healthy aorta

To compare strain in the AAA with that in healthy portions of the aorta, we conducted strain analysis for Patient 3 using uncropped images that included the healthy proximal and distal sections of the aorta above and below the AAA region, as shown in Table S2. Comparing contour plots from Table 2 in the manuscript and Table S2 in the Supplementary Material clearly indicates higher strain in the healthy aorta regions, located proximally and distally to the AAA, than in the AAA. For Patient 3, the peak strain of 17.77% in the healthy aorta is significantly higher than the peak strain of 7.49% in the AAA region. This is consistent with the results for 99th percentile strain where the 99th percentile strain in the AAA wall is 4.46%, while in the healthy aorta, it reaches 9.21%. Interpretation of these results in the terms of healthy aorta and AAA tissue mechanical properties is beyond the scope of this study.

**Table S2.** Kinematic analysis results of the AAA and healthy aorta regions for Patient 3, including geometry, displacement vector plots, normal displacement contour plots, and strain contour plots. The contour limits, including the 99th percentile of displacement magnitude $U_\circ$, 99th percentile of normal displacement $u_\circ$, and 99th percentile of strain $\epsilon_\circ$, are reported.

<table>
<tr><td>Geometry</td><td>Displacement vector (mm)<br>$-U_\circ$ 0 $+U_\circ$<br>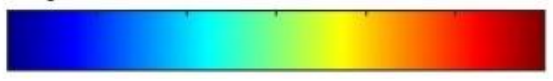</td><td>Normal displacement (mm)<br>$-u_\circ$ 0 $+u_\circ$<br>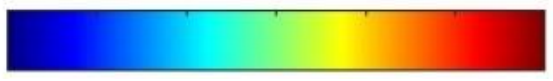</td><td>Strain<br>$-\epsilon_\circ$ 0 $+\epsilon_\circ$<br>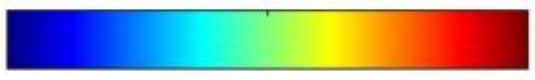</td></tr>
<tr><td>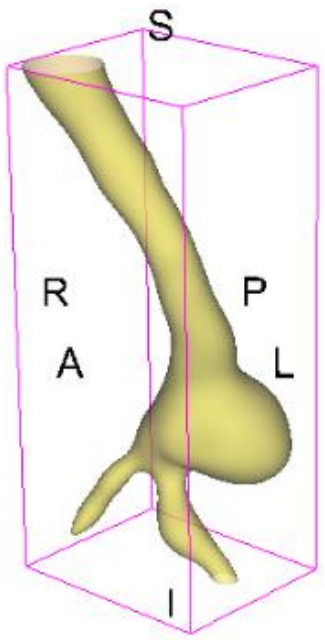</td><td>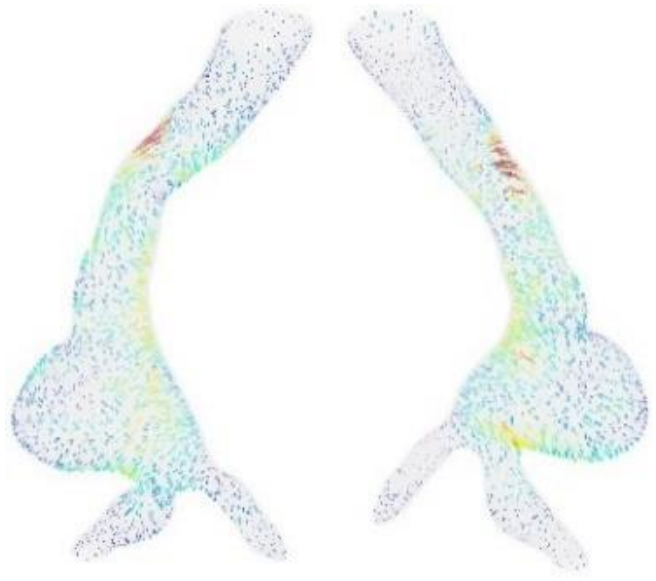<br>$U_\circ = 1.13$ mm</td><td>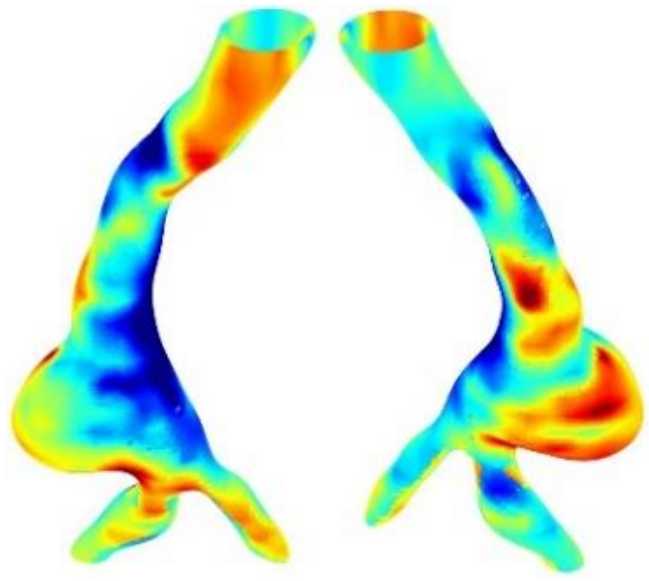<br>$u_\circ = 0.88$ mm</td><td>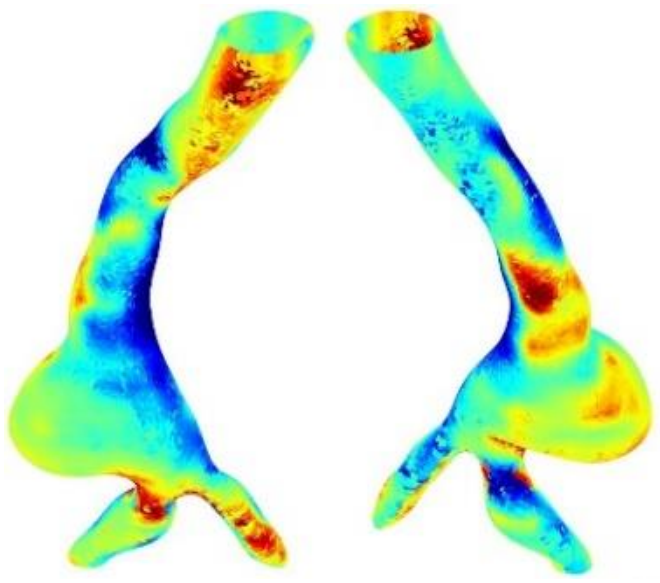<br>$\epsilon_\circ = 9.21$ %</td></tr>
</table>